\documentclass[lettersize,journal]{IEEEtran}
\usepackage{amsmath,amsfonts}

\usepackage{algorithmic}
\usepackage{algorithm}
\usepackage{array}
\usepackage[caption=false,font=normalsize,labelfont=sf,textfont=sf]{subfig}
\usepackage{textcomp}
\usepackage{stfloats}
\usepackage{url}
\usepackage{verbatim}
\usepackage{graphicx}
\usepackage{cite}
\usepackage{makecell}
\usepackage{booktabs}
\usepackage{hyperref}
\usepackage[dvipsnames]{xcolor}
\usepackage{bm} 
\usepackage{enumitem}

\usepackage{xcolor}
\usepackage{colortbl}
\usepackage{amsmath}
\usepackage{multirow}
\usepackage{amssymb,pifont}

\newcommand{\cmark}{\ding{51}} 
\newcommand{\xmark}{\ding{55}} 

\allowdisplaybreaks[3]
\begin{document}

\title{Qubit-Efficient Quantum Annealing for Stochastic Unit Commitment  }

\author{Wei Hong,~\IEEEmembership{Student Member,~IEEE}, Wangkun Xu,~\IEEEmembership{Member,~IEEE}, and Fei Teng,~\IEEEmembership{Senior Member,~IEEE}
}

\markboth{Submitted to IEEE}%
{Shell \MakeLowercase{\textit{et al.}}: A Sample Article Using IEEEtran.cls for IEEE Journals}

\maketitle

\begin{abstract}
Stochastic Unit Commitment (SUC) has been proposed to manage the uncertainties driven by renewable integration, but it leads to significant computational complexity. When accelerated by Benders Decomposition (BD), the master problem becomes binary integer programming, which is still NP-hard and computationally demanding for classical methods. Quantum Annealing (QA), known for efficiently solving Quadratic Unconstrained Binary Optimization (QUBO) problems, presents a potential solution. However, existing quantum algorithms rely on slack variables to handle linear binary inequality constraints, leading to increased qubit consumption and reduced computational efficiency. To solve the problem, this paper introduces the Powell-Hestenes-Rockafellar Augmented Lagrangian Multiplier (PHR-ALM) method to eliminate the need for slack variables, making qubit consumption independent of the increasing number of Benders cuts. To further reduce the qubit overhead, quantum ADMM is applied to break large-scale SUC into smaller blocks for sequential solutions, which does not scale with the number of generators.  Finally, the simulation results on both 4-generator and the IEEE bus-118 systems demonstrate the feasibility and scalability of the proposed algorithm, indicating its superior qubit and runtime efficiency over classical and baseline quantum approaches on the D-Wave QPU platform.

\end{abstract}


\section{Introduction}

\subsection{Unit Commitment Problem}

\IEEEPARstart{A}{S} the penetration of renewable energy continues to increase, how to efficiently handle the uncertainties associated with these resources in the system operation remains an open question \cite{bellizio2023transition}. To ensure secure and reliable operation, the system operator implements unit commitment (UC) in day ahead to decide the online status of the generators that minimizes production costs while ensuring supply and demand balance, given energy forecasts and various operational constraints. The problem can be mathematically formulated as a mixed-integer programming (MIP) problem \cite{Nasri2016NetworkConstrainedAC}, which is NP-hard \cite{Colonetti2020CombiningLR} and poses significant challenges to classic computational methods. 

Depending on different treatments of uncertainties, UC can be classified into deterministic, robust, and stochastic UC. In deterministic UC models, parameters such as load demand and renewable generation are assumed to be precisely known, resulting in a relatively smaller scale problem. Unlike deterministic models, the robust \cite{Velloso2020TwoStageRUC,Xiong2023MultistageRD,Lorca2017MultistageRUC} and stochastic \cite{Zhang2017ChanceConstrainedTUC,Huang2014TwostageSUC,Asensio2016StochasticUC} UC problems explicitly account for uncertainties as random variables \cite{teng2016stochastic}. In SUC, uncertainty is modeled using probabilistic methods by generating a representative set of scenarios through sampling. However, achieving a representative and feasible solution may require a large number of scenarios, significantly increasing computational complexity. Multi-block Alternating Direction Method of Multipliers (ADMM) and Benders Decomposition (BD) are two algorithms that can efficiently reduce the problem size and allow parallel processing. However, dealing with large numbers of integer variables and scenarios is still challenging for a real-time application.

\subsection{Quantum Computing for Power System Operations}

Quantum computing (QC), an emerging computational paradigm grounded in quantum mechanics \cite{Nielsen2010QC}, provides a novel architecture with the potential to overcome current computational bottlenecks. Unlike classical computing, where the fundamental unit is the bit, QC uses quantum bits, or qubits, to store information. The essence of QC involves exploiting quantum entanglement and superposition to create and manipulate complex multi-eigenstate superpositions \cite{Nielsen2010QC}. This capability allows quantum computers to process multiple eigenstates in parallel. Currently, QC has been developed under two primary computational models: the \emph{gate-based model} and the \emph{quantum annealing (QA) based model} \cite{Morstyn2024OpportunitiesQC}. The gate-based model constructs quantum circuits, such as quantum approximate optimization algorithm (QAOQ), uses quantum gates to manipulate qubit states, thereby evolving the probabilities of all eigenstates to achieve specific computational goals. On the other hand, QA utilizes the phenomenon of quantum tunneling to identify the lowest energy state of the system, known as the ground state \cite{Pastorello2019QuantumAL}. In this ground state, the spin orientations of qubits can be interpreted as the global optimal solution to a Quadratic Unconstrained Binary Optimization (QUBO).

In power systems, the potential of QC has been explored across problems of increasing structural complexity. Early efforts focused on linear system applications, where the Harrow-Hassidim-Lloyd algorithm provides exponential acceleration for solving linear problems such as DC optimal power flow \cite{Amani2023QuantumenhancedDC} and fast decoupled load flow \cite{Feng2021QuantumPF}, compared to classical methods. Beyond deterministic linear models, QC has been extended to uncertainty quantification. For example, \cite{Nikmehr2023QuantuminspiredPS} proposes a Quantum Monte Carlo Simulation (MCS) method based on Quantum Amplitude Estimation for reliability assessment under uncertainty, demonstrating quadratic speedup over classical MCS.

More recently, QC has been applied to mixed-integer optimization problems in power systems, many of which can be reformulated into QUBO form. For instance, QAOA has been used for optimal electric vehicle charging \cite{Kimleang2023LeveragingKQAOA}, and QA has been employed to solve combinatorial optimal power flow \cite{Morstyn2023AnnealingbasedQC} problems using D-Wave quantum processors. Building upon these developments, QC offers opportunities for more complex mixed-integer problems such as unit commitment (UC)\cite{Mahroo2023LearningIQ, Nikmehr2022QuantumDUC, Feng2023NovelRUC, Halffmann2022QCApproach}.

Given that QC primarily targets integer programming, the BD framework is appealing because the master problem explicitly includes all first-stage integer variables in a two-stage problem such as SUC. Table \ref{tab:comparison_uc} provides a comparison of recent BD-based quantum-assisted formulations for power system applications, highlighting their treatment of slack variables, additional qubit requirements, and convergence behavior. \emph{A key challenge lies in the efficient management of the increasing number of Benders cut constraints as the algorithm iterates.} In detail, slack variables are often used to convert the resulting inequality constraints into equalities \cite{Mahroo2023LearningIQ, Nikmehr2022QuantumDUC, Feng2023NovelRUC, Wang2023QuantumAIS, Leenders2024IntegratingQCC,FU2023CoordinatedPR, Halffmann2022QCApproach, barrass2025leveraging, Paterakis2023hybridQC, Ling2025hybridQA, gao2025distributed}.  However, enabling QPUs to handle these slack variables requires a substantial allocation of qubits for binary encoding (a technique to convert continuous variables into binary variables), whose qubit occupation increases linearly with the Benders' iterations. This leads to significant qubit overhead, reduced computational efficiency, or even causes chain break, a catastrophic failure mode of QC. Recent hybrid quantum UC studies\cite{barrass2025leveraging,Ling2025hybridQA,gao2025distributed} have attempted to reduce the effect of slack-variable or cut-related encodings, mainly by lowering their encoding cost or reducing the size of quantum subproblems after auxiliary variables have been introduced. However, as the slack variable always exists, the qubit overhead caused by these variables remains a key limitation for scalable QA-based UC formulations.

\begin{table*}[t]
\centering
\caption{Recent BD-based quantum-assisted approaches for power system applications.}
\label{tab:comparison_uc}
\footnotesize
{
\small
\setlength{\tabcolsep}{4pt}

\begin{tabular}{l c c c c c c c}
\hline
\hline
\multirow{2}{*}{\textbf{Reference}}
& \multirow{2}{*}{\textbf{\makecell{Quantum\\Algorithm}}}
& \multirow{2}{*}{\textbf{Problem}}
& \multirow{2}{*}{\textbf{\makecell{Binary-encoded\\Slack Variables}}}
& \multicolumn{2}{c}{\textbf{Source of Additional Qubits}}
& \multirow{2}{*}{\textbf{\makecell{Component(s) \\with Convergence\\ Guaranteed}}} \\
\cline{5-6}
& & & 
& \textbf{\makecell{Benders\\Cuts} }
& \textbf{\makecell{Power System\\Constraints}}
& \\
\hline
\cite{Leenders2024IntegratingQCC} & QA & Multi-energy System  & \cmark & \cmark & \cmark & \xmark \\
\cite{FU2023CoordinatedPR} & QA & Post-disaster Restoration  & \cmark & \cmark & \cmark & \xmark \\
\cite{zhao2023optimal} & QA & Energy Management  & \cmark & \cmark & \cmark & \xmark \\
\cite{wei2024hybrid} & QA & Federated Learning Scheduling  & \cmark & \cmark & \cmark & BD \\
\hline
\hline
\cite{barrass2025leveraging} & QA & UC  & \cmark & \cmark & \cmark & \xmark \\
\cite{Paterakis2023hybridQC} & QA & UC  & \cmark & \cmark & \cmark & BD \\
\cite{Ling2025hybridQA} & QA & UC  & \cmark & \cmark & \cmark & QA \\
\cite{gao2025distributed} & QAOA/QA & Security-constrained UC  & \cmark & \cmark & \cmark & BD and QAOA \\
This work & QA & SUC  & \xmark & \xmark & \xmark & BD and QA \\
\hline
\hline
\end{tabular}
}

\end{table*}

\subsection{Contributions}

This paper studies the efficient use of QA for solving SUC within the BD framework. Combining the Powell-Hestenes-Rockafellar Augmented Lagrangian Multiplier (PHR-ALM) and Direct-extended ADMM (D-ADMM), Quantum-based PHR ADMM (QPHR-ADMM) is proposed. The qubit occupation of the proposed method does not scale with the number of Benders iteration nor the number of generators, leaving it scalability for large-scale power system testbed with limited quantum resources.
Detailed contributions of this paper are summarized as:

\begin{enumerate}[leftmargin=*]
    \item{The PHR-ALM method is proposed to \emph{completely eliminate} the need of slack variables for inequality constraint embedding within the BD framework, which breaks down the linear dependency between the number of qubits and the number of BD iterations.}

    \item{Within each BD iteration, a D-ADMM algorithm is designed to decomposes large-scale SUC into smaller blocks, enabling sequential solves and eliminating the linear scaling with the number of generators; the resulting complexity depends primarily on the time horizon and the binary-encoding precision.}

    \item For both PHR-ALM and D-ADMM, we provide analytical derivations of their QUBO and Hamiltonian representations, making them directly implementable with QA algorithms.
\end{enumerate}

By the above contributions, this work is able to extend the previous application of QA for large-scale UC problem. Notably, Prior works \cite{Halffmann2022QCApproach,Paterakis2023hybridQC,Ling2025hybridQA} on D-Wave QPUs are usually restricted to systems with maximum 15 units and often do not consider a full 24-hour scheduling horizon. By contrast, through various case studies on synthetic, 4-generator, and realistic IEEE bus-118 test cases, we demonstrate the qubit and computational time efficiencies as well as the scalability of the proposed algorithm, compared to baseline quantum methods.

To sum up, with advancements in quantum technology, QA is expected to surpass classical computation for solving SUC. More broadly, this paper demonstrates that even with limited quantum resources, the potential of QC can be effectively harnessed through rigorous mathematical reformulations, accelerating  the deployment of quantum-integrated optimization for power system applications. The structure of this paper is organized as follows. Section \ref{section:II} describes the scenario-based SUC problem, solved by BD framework, along with the overview of QC technology. Particularly, we discuss why QA is chosen, compared to gate-based method, such as QAOA. Section \ref{section:III} details the design process of the QPHR-ADMM algorithm. Case studies are presented in Section \ref{section:IV}. Finally, Section \ref{section:V} concludes the paper.

\section{Preliminaries}
\label{section:II}

\subsection{Stochastic Unit Commitment}\label{section:II A}
\subsubsection{SUC Problem Formulation}

To model the uncertainty of wind power generation on the supply side and the uncertainty of electricity consumption on the demand side, a two-stage scenario-based SUC model can be applied. The first stage involves making day-ahead UC decisions that satisfy all unit constraints, primarily determining the on/off status of thermal generators. The second stage involves the economic dispatch of thermal generators under each uncertain scenario, while ensuring compliance with operation constraints. 
A complete formulation of the scenario-based SUC is considered as follows\cite{Pablo2009UncertaintyMUC,Ignacio2017AnEfficientRS,Canan2016AnImprovedSUC}.
\begin{subequations}\label{eq:1}
\setlength\abovedisplayskip{1pt}
\setlength\belowdisplayskip{5pt}
\begin{align}
\min &\sum\limits_{g=1}^{N} \sum\limits_{t=1}^{T} C_{g}^{cons} u_{g,t}+\sum\limits_{h=1}^{K} \pi\left(\xi_{h}\right) \Bigg\{ \sum\limits_ { g = 1 } ^ { N } \sum\limits_ { t = 1 } ^ { T } \bigg[ C_{g}^{quad}\left(P_{h, g, t}^{G}\right)^{2} \nonumber\\ 
&+ C_{g}^{prim } P_{h, g, t}^{G} \bigg]+\sum_{t=1}^{T}C^{shed
 } P_{h, t}^{shed
 } \Bigg\} \label{eq:1a}
\end{align}
\begin{align}
&\text{subject to:} \nonumber\\
& u_{g,t} P_{g}^{min}
\le P_{h,g,t}^{G}
\le u_{g,t} P_{g}^{max}
\quad \forall h,g,t \label{eq:1b}\\
& P_{g}^{rd}
\le P_{h,g,t+1}^{G}-P_{h,g,t}^{G}
\le P_{g}^{ru}
\quad \forall h,g,t \label{eq:1c}\\
& \sum_{g=1}^{N} P_{h,g,t}^{G}
+ P_{h,t}^{shed}
+ P_{h,t}^{Wind}
= D_{h,t}^{Load}
\quad \forall h,t \label{eq:1d}\\[3pt]
&
\sum_{\varsigma=t}^{t+T_g^{U}-1} \!\!\!\!u_{g,\varsigma}\!\!
\;\ge\;\!\!
T_g^{U}\,\!\!\bigl(u_{g,t}\!-\!u_{g,t-1}\bigr)\!\!\!
\quad \forall g,\;\!\! t\!=\!2,\!\dots\!,T\!-\!T_g^{U}\!\!+1\!\!\!
 \label{eq:1e}\\[3pt]
&
\sum_{\varsigma=t}^{t+T_g^{D}-1} \!\!\!\!\!\bigl(1\!-\!u_{g,\varsigma}\bigr)
\;\!\!\ge\;\!\!
T_g^{D}\,\!\!\bigl(u_{g,t-1}\!-\!u_{g,t}\bigr)
\!\!\!\!\quad \forall g,\;\!\! t\!=\!2,\!\dots\!,T\!-\!T_g^{D}\!\!+\!1
\label{eq:1f}\\[3pt]
& u_{g,t} \in \{0,1\}
\quad \forall g,t \label{eq:1g}\\
& P_{h,t}^{shed} \ge 0
\quad \forall h,t \label{eq:1h}
\end{align}
\end{subequations}

In detail, during an operation horizon of $T$ periods, the decision variables are categorized into two groups. The first-stage decision variables are binary variables $u_{{g},{t}}$ determining the on/off states for the $g$-th unit in time period $t$. Since these first-stage decisions pertain to day-ahead scheduling, they are independent of realizations of uncertainties. Subsequently, when wait-and-see decisions are made, uncertainties in the power system are realized in terms of the output of wind farms on the supply side and the load on the demand side. The corresponding random variables are defined as $P_{{h},{t}}^{Wind}$ and $D_{{h},{t}}^{Load}$, respectively, for period $t$ for scenario $h$ in the scenario set $\Omega$ with probability $\pi \left ( \xi_{h} \right )$. Due to uncertainties in both supply and demand in the second stage, load shedding $P_{{h},{t}}^{shed}$ is introduced to ensure the feasibility. Moreover, $C_{g}^{quad}$, $C_{g}^{prim}$ and $C_{g}^{cons}$ are cost coefficients. $C^{shed}$ is the penalty cost of load shedding. $P_{g}^{min}$ and $P_{g}^{max}$ donate the minimum and maximum power output. $P_{g}^{rd}$ and $P_{g}^{ru}$ are ramp-up and ramp-down rate limits. $T_{g}^{U}$ and $T_{g}^{D}$ denote the minimum up-time and down-time durations, respectively.
The objective in SUC minimizes the total cost, consisting of constant production costs and the quadratic fuel costs of thermal units in (\ref{eq:1a}).  At last, (\ref{eq:1d}) represents the power balance constraints for any given period in any scenario.

\subsubsection{Solution Algorithm Based on Benders Decomposition}
The BD method has been widely applied in large-scale power system optimization such as the SUC to reduce its computational burden\cite{Gao2022HybridQC,Leenders2024IntegratingQCC,Roald2023PowerSO}. 
It decouples complex, multi-constraint problems into small-scale subproblems that can be solved in parallel. 
In (\ref{eq:1b}), the first-stage variables are incorporated into the second-stage constraints, leading to a coupling between the two stages. To decouple the two stages, BD divides the SUC problem (\ref{eq:1a})-(\ref{eq:1h}) into two sections, i.e., a mixed-integer master problem and \emph{independent} scenario-based quadratic subproblems. The master problem captures the first-stage commitment decisions $u_{g, t}$. At iteration $k$, the master problem is written as,
\begin{subequations}\label{eq:2}
\setlength\abovedisplayskip{1pt}
\setlength\belowdisplayskip{5pt}
\begin{align}
\min_{u_{g,t}^{k}, \Upsilon^{k}}
&\sum_{g=1}^{N} \sum_{t=1}^{T} C_{g}^{cons} u_{g,t}^{k}
+ \Upsilon^{k} \label{eq:2a}
\end{align}
\begin{align}
&\text { subject to: } \nonumber\\
&\Upsilon^{k} \geq \alpha^{lower }\label{eq:2b}\\
&\Upsilon^{k} \geq \sum_{h=1}^{K} \pi\left(\xi_{h}\right)\left\{\sum_{g=1}^{N} \sum_{t=1}^{T}\bigg[C_{g}^{quad}\left(P_{h, g, t}^{G, l}\right)^{2}\right. \nonumber\\
& \hspace{2cm} \left.+C_{g}^{prim } P_{h, g, t}^{G, l}\bigg]+\sum_{t=1}^{T}C^{shed} P_{h, t}^{shed, l}\right\} \nonumber\\
&\quad\!\!+\sum_{g=1}^{N}\!\sum_{t=1}^{T} \limits\!\sum_{h=1}^{K} \theta_{g, t}^{l}\left(\xi_{h}\right)\!\left[u_{g,t}^{k}-u_{g,t}^{l}\right] \!\!\!\quad \forall l=0, \ldots, k\!-\!1\label{eq:2c}\\[3pt]
&
\sum_{\varsigma=t}^{t+T_g^{U}-1} \!\!\!\!u_{g,\varsigma}\!\!
\;\ge\;\!\!
T_g^{U}\,\!\!\bigl(u_{g,t}\!-\!u_{g,t-1}\bigr)\!\!\!
\quad \forall g,\;\!\! t\!=\!2,\!\dots\!,T\!-\!T_g^{U}\!\!+1\!\!\!
\label{eq:2d}\\[3pt]
&
\sum_{\varsigma=t}^{t+T_g^{D}-1} \!\!\!\!\!\bigl(1\!-\!u_{g,\varsigma}\bigr)
\;\!\!\ge\;\!\!
T_g^{D}\,\!\!\bigl(u_{g,t-1}\!-\!u_{g,t}\bigr)
\!\!\!\!\quad \forall g,\;\!\! t\!=\!2,\!\dots\!,T\!-\!T_g^{D}\!\!+\!1
 \label{eq:2e}\\[3pt]
&u_{g, t}^{k} \in\{0,1\} \quad \forall g, t \label{eq:2f}
\end{align}
\end{subequations}
where $\theta_{g, t}^{l}\left(\xi_{h}\right)$s are the optimal dual variables; $P_{h,g,t}^{G,l}$ and $P_{h,t}^{shed,l}$ are the optimal primal variables. Both are introduced by the subproblems. Note that, due to the introduction of a continuous Benders lower bound (LB), $\Upsilon $, the master problem is characterized as an MIP problem rather than a pure integer program.

The subproblem for scenario $h$ is given as (\ref{eq:3a})-(\ref{eq:3f}) below. As each scenario is independent, the subproblem is decoupled and can be treated separately. The main objective is to strategize the optimal generation levels and corresponding load shedding for each scenario.
\begin{subequations}\label{eq:3}
\setlength\abovedisplayskip{1pt}
\setlength\belowdisplayskip{5pt}
\begin{align}
\min _{u_{g, t}^{fixed }, P_{h, t}^{G, k}, P_{h, t}^{p, k}, P_{h, t}^{p_{t}, k}} & \pi\left(\xi_{h}\right)\Bigg\{\sum_{g=1}^{N} \sum_{t=1}^{T}\bigg[C_{g}^{q u a d}\left(P_{h, g, t}^{G, k}\right)^{2} \nonumber\\
 +C_{g}^{p r i m} P_{h, g, t}^{G, k} &\bigg]+\sum_{t=1}^{T}C^{shed} P_{h, t}^{shed, k}\Bigg\}\label{eq:3a}
\end{align}
\begin{align}
&\text { subject to: } \nonumber\\
&u_{g, t}^{fixed} P_{g}^{\min } \leq P_{h, g, t}^{G, k} \leq u_{g, t}^{fixed} P_{g}^{\max } \quad \forall h, g, t \label{eq:3b}\\
&P_{g}^{r d} \leq P_{h, g, t+1}^{G, k}-P_{h, g, t}^{G, k} \leq P_{g}^{r u} \quad \forall h, g, t \label{eq:3c}\\
&\sum_{g=1}^{N} P_{h, g, t}^{G, k}+P_{h, t}^{shed, k}+P_{h, t}^{W i n d}=D_{h, t}^{L o a d} \quad \forall h, g, t \label{eq:3d}\\
&u_{g, t}^{fixed }=u_{g, t}^{k}: \theta_{g, t}^{k}\left(\xi_{h}\right) \quad \forall h, g, t\label{eq:3e}\\
&P_{h, t}^{shed, k} \geq 0\quad \forall h, t \label{eq:3f}
\end{align}
\end{subequations}
In each subproblem, $u_{g,t}^{fixed}$ is a newly introduced auxiliary continuous variable and $u_{g,t}^{k}$ is the optimal decision variable of the master problem. As a result, \eqref{eq:3} results in an upper bound (UB) of \eqref{eq:1}. Meanwhile, due to the convexity of \eqref{eq:3}, the dual variable $\theta_{g, t}^{l}\left(\xi_{h}\right)$s in \eqref{eq:3e} defines the sensitivity of subproblem's objective. After collecting all sensitivities, a new Benders cut (\ref{eq:2c}) can be added in the next iteration and the master problem \eqref{eq:2} becomes a LB to \eqref{eq:1}. Moreover, as subproblem is a continuous quadratic programming, it can be efficiently solved on CPUs in parallel.

The BD method decouples the first-stage and second-stage operations, significantly reducing the problem size. It is noteworthy that all integer variables are now included in the master problem where QC holds the potential to further improve its computational efficiency.

\subsection{Quantum Computing
} \label{section:II B}

This section provides the QC background and the conversion of classical QUBO problems into QC-ready forms. Then we introduce QA based on Dwave QPUs to solve QUBO problems as Fig. \ref{QA_framework} showing, and discuss why QA is chosen over QAOA for solving SUC master problem \eqref{eq:2}.

\subsubsection{QUBO Problem and Hamiltonian Representation}
QUBO, an NP-hard problem, is one of the most widely studied optimization problems in QC. In this model, the solution to the optimization problem is composed of a series of binary variables whose value is $0$ or $1$, and these variables correspond exactly to the off or on state of units in \eqref{eq:2}. The general expression of the QUBO problem is given as 

\begin{equation}
\setlength\abovedisplayskip{1pt}
\setlength\belowdisplayskip{5pt}
\label{eq:4}
\min \sum_{i \in \aleph } \sum_{j \in \aleph , i \neq j} B_{i j} \tau_{i} \tau_{j}+\sum_{i \in \aleph } c_{i} \tau_{i}
\end{equation}
where $\aleph =\{1, \ldots, \mathfrak{n} \}$ denote the index set of binary variables. The variables include $\tau_{i}, \tau_{j} \in\{0,1\}$, and $B_{i j}, c_{i} \in \mathbb{R}$ are the coefficients of quadratic and linear terms, respectively. To enable the execution of QUBO problems on quantum hardware, it must be transformed into Ising-Lenz model. In quantum mechanics, the binary variables represent particle spin directions, where spin-up $\left | \uparrow  \right \rangle=\begin{bmatrix} 1&0 \end{bmatrix}^{\mathrm{T}}$ and spin-down $\left | \downarrow  \right \rangle = \begin{bmatrix} 0&1 \end{bmatrix}^{\mathrm{T}}$ correspond to $\upsilon=+1$ and $\upsilon=-1$, respectively \cite{Nielsen2010QC}. These states are conventionally mapped to $\left | 0  \right \rangle$ and $\left | 1  \right \rangle$ in QC, leading to the variable substitution $\tau=(1-\upsilon) / 2$. Based on this relation, the Ising model can be derived from (\ref{eq:4}) as
\begin{equation}\label{eq:5}
\setlength\abovedisplayskip{1pt}
\setlength\belowdisplayskip{5pt}
\begin{split}
   \min \sum_{i \in \aleph } \sum_{j \in \aleph , i \neq j} Q_{i j} \upsilon_{i} \upsilon_{j}+\sum_{i \in \aleph } p_{i} \upsilon_{i}+  \text{Const.}\quad \text{with} \\
   Q_{i j}=\frac{B_{i j}}{4}, \quad p_{i}=-\sum_{j \in \aleph , i \neq j}\left(\frac{B_{i j}+B_{j i}}{4}\right)-\frac{c_{i}}{2}
\end{split}
\end{equation}

While the Ising model encodes the cost function, quantum state evolution in QA and QAOA is governed by the Hamiltonian. Hence, as shown by Fig. \ref{QA_framework}, constructing a suitable Hamiltonian representation from Ising model is essential. This is achieved by mapping spin eigenstates $\upsilon$ to their eigenvalues using the Pauli-$z$ operator $\boldsymbol{\sigma}^z$, allowing the Hamiltonian form of the QUBO problem to be obtained by substitution in (\ref{eq:5}). Based on (\ref{eq:5}) and (\ref{eq:6}), we can construct the graphs of the QUBO problem.

\begin{equation}\label{eq:6}
\setlength\abovedisplayskip{1pt}
\setlength\belowdisplayskip{5pt}
\begin{aligned}
\boldsymbol{H}_{Q U B O} & =\sum_{i \in \aleph } \sum_{j \in \aleph , i \neq j} Q_{i j} \boldsymbol{\sigma}_{i}^{z} \otimes \boldsymbol{\sigma}_{j}^{z}+\sum_{i \in \aleph } p_{i} \boldsymbol{\sigma}_{i}^{z},\; \text{with} \\
\boldsymbol{\sigma}^{z} & =\left[\begin{array}{cc}
1 & 0 \\
0 & -1
\end{array}\right]
\end{aligned}
\end{equation}

\subsubsection{Quantum Annealer}
QA is a quantum algorithm designed for quantum annealers, which leverages quantum tunneling, induced by quantum fluctuations, to escape suboptimal local minima and search for the global minimum of an objective function across a vast solution space \cite{Pastorello2019QuantumAL}. By enabling direct transitions through energy barriers, QA can efficiently reach the global optimum, often outperforming classical algorithms in solving QUBO problems.

The design of QA is based on the principle of adiabatic evolution. This principle states that if the system is initialized in the ground state of a known Hamiltonian, and the evolution is sufficiently slow with no energy level crossings, the system will remain in its instantaneous ground state throughout \cite{Morstyn2023AnnealingbasedQC,Albash2018AdiabaticQC}. When the final Hamiltonian encodes the QUBO objective function via the Ising model, the resulting ground state corresponds to the optimal solution. However, if the energy barrier is too wide, the system may still evolve to a local minimum. The mathematical expression of this process is as
\begin{equation}\label{eq:7}
\setlength\abovedisplayskip{1pt}
\setlength\belowdisplayskip{5pt}
\boldsymbol{H}_{mixed}(t)=[1-s(t)] \boldsymbol{H}_{I}+s(t) \boldsymbol{H}_{F}
\end{equation}
where $\boldsymbol{H}_{I}$ is an initial Hamiltonian, $\boldsymbol{H}_{F}$ is a final Hamiltonian that equals the objective Hamiltonian, and $s(t)$ is referred to as the normalized evolution time.

\begin{figure}[!t]
\centering
\includegraphics[width=3.5 in]{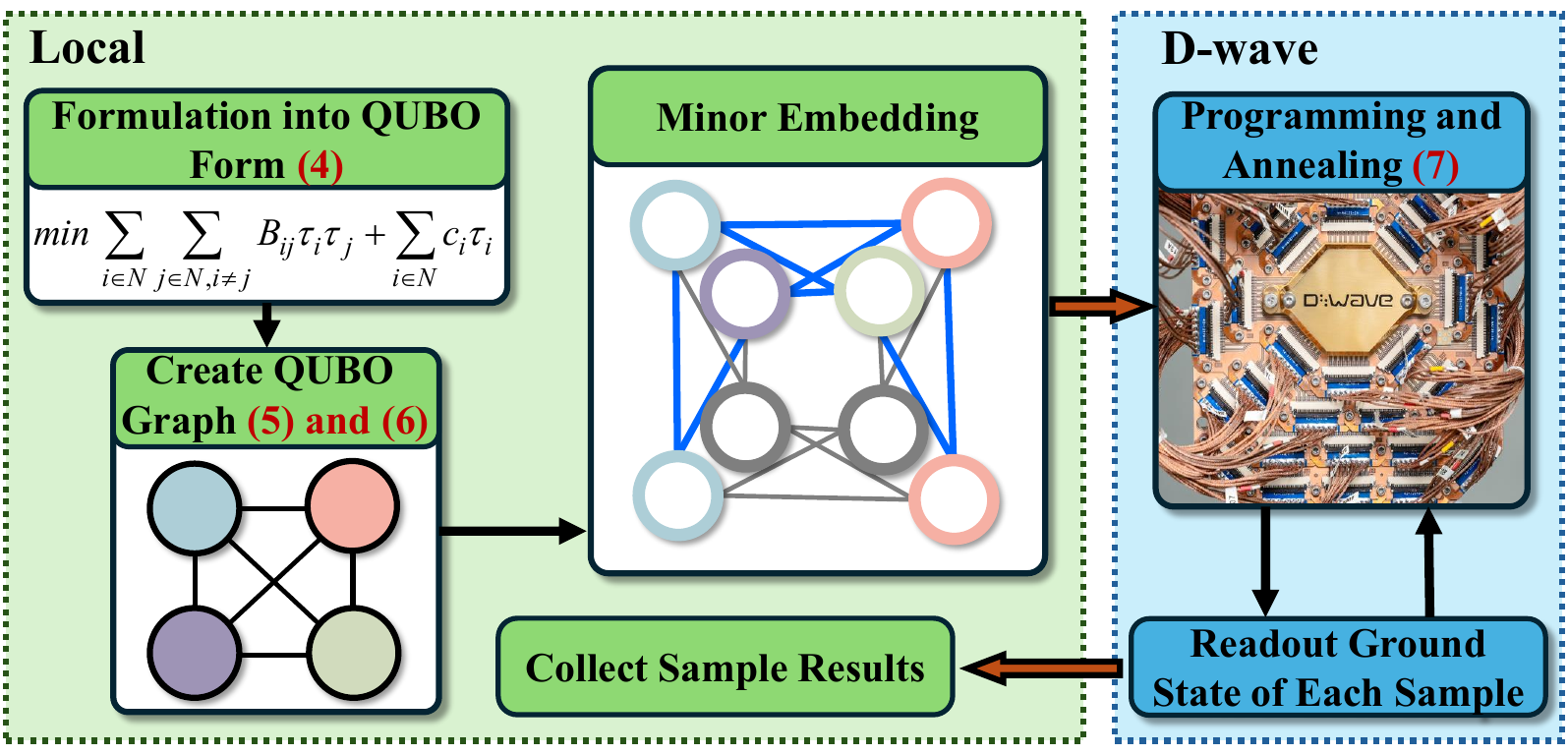}
\caption{The framework of QA using Dwave QPU to solve QUBO problems.}
\label{QA_framework}
\vspace{-1em}
\end{figure}

Overall, once the objective is formulated as a QUBO problem, a minor embedding is applied to map the QUBO graph onto the QPU’s sparse hardware connectivity by chaining multiple physical qubits per logical variable, and then the QPU is annealing to the lowest-energy eigenstate. The described QA procedure (\ref{eq:7}) is currently supported by D-Wave quantum annealer as illustrated in Fig. \ref{QA_framework}. In D-Wave 2000Q, qubits are implemented as tiny metal loops arranged in a Chimera lattice, where each unit cell contains 8 interconnected qubits.

\subsubsection{Digital Quantum Computers}\label{sec:qaoa_into}

Digital (gate-based) quantum computers provide a general-purpose quantum computing platform based on quantum gates and circuit-based operations. A variety of quantum algorithms can be implemented on such platforms for combinatorial optimization problems.

QAOA is one representative variational algorithm designed for gate-based quantum computers, and it operates in a hybrid classical–quantum framework. While not strictly adiabatic, QAOA draws inspiration from the adiabatic principle by constructing a depth-$\mathcal{D}$ ansatz via alternating applications of $\boldsymbol{H}_{I}$ and $\boldsymbol{H}_{F}$ \cite{Farhi2014AQuantumAOA,Blekos2024AreviewQAOA}. As a first step, an initial uniform state $\left| \phi_0 \right\rangle$ is prepared by applying Hadamard gates $\bm{\mathcal{H}}$, resulting in a ground state of the initial Hamiltonian $\boldsymbol{H}_{I}$, as shown in \eqref{eq:pat1}.
\vspace{0.1em}
\begin{equation}\label{eq:pat1}
\setlength\abovedisplayskip{1pt}
\setlength\belowdisplayskip{0pt}
\left| \phi_0 \right\rangle = \bm{\mathcal{H}}^{\otimes \aleph} \left| 0 \right\rangle^{\otimes \aleph} = \bigotimes_{i=1}^{\aleph} \frac{\left| 0 \right\rangle + \left| 1 \right\rangle}{\sqrt{2}}= \left| + \right\rangle^{\otimes \aleph}
\end{equation}
After initialization, a depth-\(d\) parameterized ansatz is constructed using alternating unitaries $\boldsymbol{U}_{I}^{d}$ and $\boldsymbol{U}_{F}^{d}$, derived from $\boldsymbol{H}_{I}$ and $\boldsymbol{H}_{F}$, respectively. These unitaries are sequentially applied and parameterized by variational angles $\boldsymbol{\beta} = (\beta_1, \ldots, \beta_d)$ and $\boldsymbol{\gamma} = (\gamma_1, \ldots, \gamma_d)$, to simulate a discretized adiabatic evolution. The resulting quantum state is given in \eqref{eq:pat2}, 
\vspace{0.1em}

\begin{equation}\label{eq:pat2}
\setlength\abovedisplayskip{-8pt}
\setlength\belowdisplayskip{0pt}
\!\!\left| \phi(\boldsymbol{\beta}, \boldsymbol{\gamma}) \right\rangle 
= \prod_{d=1}^{\mathcal{D}} \boldsymbol{U}_{I}^{d} \boldsymbol{U}_{F}^{d} \left| \phi_0 \right\rangle 
= \prod_{d=1}^{\mathcal{D}} e^{-i \beta_d \boldsymbol{H}_{I}} e^{-i \gamma_d \boldsymbol{H}_{F}} \left| \phi_0 \right\rangle
\end{equation}

Finally, the parameters are updated through classical optimization to minimize the expectation value of the objective Hamiltonian, 
$\langle \phi(\boldsymbol{\beta}, \boldsymbol{\gamma}) | \boldsymbol{H}_{F} | \phi(\boldsymbol{\beta}, \boldsymbol{\gamma}) \rangle$, 
which is estimated by measuring the evolved quantum state.

\subsubsection{Quantum Annealers vs Digital Quantum Computers}\label{sec:qaoa_into}
From a practical perspective, existing studies \cite{Kim2024DistributedQAOA,Pelofske2023QuantumAnnealingvsQAOA} indicate that the scalability of digital quantum computers for large-scale integer programming remains constrained by current hardware. First, gate-based QPU typically offer limited qubits compared to annealing-based systems. For example, the IBM Heron r2 processor provides 156 qubits, which significantly limits its scalability. Second, although QAOA can achieve asymptotic convergence of the expectation value to the ground-state energy $H_{\mathrm{min}}$ as the circuit depth increases ($\lim_{d \to \infty} \left\langle \phi(\boldsymbol{\beta}, \boldsymbol{\gamma}) \vert \boldsymbol{H}_{F} \vert \phi(\boldsymbol{\beta}, \boldsymbol{\gamma}) \right\rangle \!=\!H_{\mathrm{min}}$ \cite{Farhi2014AQuantumAOA}), practical depths are limited by noise-induced barren plateaus, where the expected gradient decays exponentially with depth and system size \cite{wang2021noise}, leading to potentially suboptimal solutions. Within the BD framework, this will result in weaker lower-bound tightening such that more iterations for convergence are required.

Therefore, although gate-based approaches such as QAOA remain promising, this study focuses on near-term feasibility under current hardware constraints. Therefore, QA is adopted as the backend in this work and a simulation comparison between QA and QAOA is also provided in Appendix \ref{section:IV A}.

\section{Proposed QPHR-ADMM Algorithm} \label{section:III}

In practice, although D-Wave QA-QPUs such as Advantage\_system 4.1 provide a large qubit capacity (up to 5,627 qubits) \cite{DWave2024QPUSpecificPP}, their usage remains limited, due to two main factors. (i) In minor embedding, the sparse connectivity of QA-QPU topology requires additional qubits for embedding QUBO problems, and (ii) The chain-breaking issue \cite{Elijah2023ComparingTG}, which compromises the integrity of the QUBO representation, can lead to incorrect solutions. These challenges hinder the scalability and effectiveness of QA in large-scale optimization.

As a result, designing a QA algorithm with minimal qubit overhead is crucial for addressing real-world problems. This section first introduces binary encoding and augmented Lagrangian function to transform \eqref{eq:2} into QUBO form and then demonstrates that a direct implementation of QA for solving \eqref{eq:2} inevitably requires an increasing number of additional qubits as the Benders' iterations progress, compounding the already significant qubit demand introduced by SUC itself. Two novel algorithms, namely PHR Augmented Lagrangian and quantum-based ADMM, are proposed to resolve the challenges with theoretical analysis on their computational complexity.

\subsection{Pre-processing for Quantization}
\label{section:III A}
As the master problem \eqref{eq:2} in Section \ref{section:II A} is a constrained MIP problem rather than in a QUBO form, QA-QPUs cannot directly construct the corresponding quantum embedding. Therefore, preprocessing techniques are presented. 

\subsubsection{Binary Encoding}

To start, the continuous variable $\Upsilon$, introduced by Benders LB \eqref{eq:2}, cannot be directly processed by the QPU. Binary encoding \cite{Zhao2021HybridQB}, expressed as $\Upsilon^{k}=\chi \sum_{j=0}^{J-1} 2^{j} u_{j}^{k}$, can be applied where $u_{j}^{k}$ are binary variables. By selecting the appropriate precision coefficients $\chi$ and precision levels $J$, the desired precision can be attained. Leveraging the $NT$ number of binary variables in the original SUC, the qubit occupation after binary encoding becomes $\underline{NT+J}$.

\subsubsection{Augmented Lagrangian Function}

In the UC formulation and the Benders decomposition framework, inequality constraints are typically handled by incorporating them into the objective function through augmented Lagrangian techniques, which will introduce extra binary encoded slack variables for execution on a QPU. This applies both to structural UC constraints, such as minimum up/down time constraints, and to the Benders cuts generated during the iterative process.

Specifically, suppose there are $\mathcal{M}$ minimum up/down time constraints, each requiring $\mathcal{P}$ qubits for binary-encoded slack variables to achieve the desired precision. This results in a fixed qubit overhead of $\mathcal{M}\mathcal{P}$. In addition, according to (\ref{eq:2c}), the number of Benders cuts grows with the iteration index $k$. Assuming the precision level of each corresponding slack variable is $\mathcal{F}$, the resulting qubit overhead increases linearly with $k$ and can be expressed as $\underline{NT + J +\mathcal{M}\mathcal{P} + k\mathcal{F}}$.

Considering the limited number of qubits in quantum computers, introducing slack variables is inefficient and unacceptable. This inefficiency serves as the primary motivation for the PHR Lagrangian algorithm proposed in this Section \ref{section:III B}. Moreover, the qubit occupation grows linearly with the number of generator $N$. This qubit overhead will be resolved by the quantum ADMM algorithm in Section \ref{section:III C}.

\subsection{Quantum PHR Augmented Lagrangian Algorithm}

\label{section:III B}

\subsubsection{Classic Formulation}
The Powell-Hestenes-Rockafellar Augmented Lagrangian Multiplier (PHR-ALM) method was originally designed for equality-constrained optimization problems. Rockfellar later extended it to handle inequality constraints by eliminating the need for slack variables in optimization problems \cite{Rockafellar1974AugmentedLM}, which is an ideal property for QA.

To start, the master problem \eqref{eq:2}, after binary encoding, can be compactly formulated as a linear binary optimization (LBO) problem.
\begin{equation}\label{eq:8}
\setlength\abovedisplayskip{1pt}
\setlength\belowdisplayskip{5pt}
\vspace{-0em}
\begin{aligned}
\min \ &f(\boldsymbol{x}) \\
\text { s.t. } &g_{i}(\boldsymbol{x}) \leq 0 \text { for } i=1, \ldots, w
\end{aligned}
\end{equation}
where $n$ represents the total number of unit states and binary-encoded variables. $\boldsymbol{x} \in\{0,1\}^{n}$ and $w$ is the total number of binary inequality constraints after some BD iterations. After introducing slack variables $\mathcal{S} _i$ to relax each inequality constraints, the augmented Lagrangian function $\mathcal{L}$ of (\ref{eq:8}) can be given as
\begin{equation}\label{eq:9}
\setlength\abovedisplayskip{1pt}
\setlength\belowdisplayskip{5pt}
\begin{aligned}
\!\mathcal{L} & =f(\boldsymbol{x})\!+\kappa(\boldsymbol{x}, \mathcal{S}, \lambda, \sigma) \\
& =f(\boldsymbol{x})\!+\!\sum_{i=1}^{w}\lambda_{i}\!\left[g_{i}(\boldsymbol{x})\!+\mathcal{S}_{i}{}^{2}\right]\!+\frac{\sigma}{2} \sum_{i=1}^{w}\left[g_{i}(\boldsymbol{x})\!+\mathcal{S}_{i}{}^{2}\right]^{2}
\end{aligned}
\end{equation}
where $\kappa$ is the augmented penalty function involving Lagrange multipliers $\lambda$ and the penalty coefficient $\sigma$.

In order to eliminate continuous slack variables $\mathcal{S}_i$s, it is necessary to derive their analytic expressions. Considering the minimization of $\mathcal{L}$ with respect to slack variables, the first-order necessary condition with respect to $\mathcal{S}$ for optimality can be expressed as
\begin{equation}\label{eq:10}
\frac{\partial \mathcal{L}}{\partial \mathcal{S}_i}
= 2\lambda_i \mathcal{S}_i + 2\sigma \mathcal{S}_i \left[g_i(\boldsymbol{x}) + \mathcal{S}_i^2\right]
= 0, \quad \forall i.
\end{equation}
Solving (\ref{eq:10}), the square form of slack variables can be obtained as
\begin{equation}\label{eq:11}
\setlength\abovedisplayskip{1pt}
\setlength\belowdisplayskip{5pt}
{\mathcal{S}_{i}{}^{2}} = \begin{cases}
\hfil{-\frac{\lambda_{i}}{\sigma}-g_{i}(\boldsymbol{x})},&\lambda_{i}+\sigma g_{i}(\boldsymbol{x}) \leq 0 \\ 
\hfil{0,}&{\lambda_{i}+\sigma g_{i}(\boldsymbol{x})>0} 
\end{cases}
\end{equation}
From (\ref{eq:11}), when $\lambda_{i}+\sigma g_{i}(\boldsymbol{x}) \leq 0$, $\mathcal{S}_{i}{}^{2}$ is non-negative, which means that the solution satisfies the $i$-th constraint $g_{i}(\boldsymbol{x}) \leq 0$. Otherwise, it indicates that the solution is not within the feasible region. Next, substituting the above results into the augmented penalty function, the corresponding PHR-based expression is given as follows,
\begin{equation}\label{eq:12}
\setlength\abovedisplayskip{1pt}
\setlength\belowdisplayskip{5pt}
{\kappa_{i}} = \begin{cases}
\hfil -\frac{\lambda_{i}{}^{2}}{2 \sigma},&\lambda_{i}+\sigma g_{i}(\boldsymbol{x}) \leq 0 \\ 
\hfil{\frac{1}{2 \sigma}\left[\left(\sigma g_{i}(\boldsymbol{x})+\lambda_{i}\right)^{2}-\lambda_{i}{}^{2}\right],}&{\lambda_{i}+\sigma g_{i}(\boldsymbol{x})>0} 
\end{cases}
\end{equation}
Combining the above two situations, the PHR-based augmented Lagrangian function can be concisely rewritten as
\begin{equation}\label{eq:13}
\setlength\abovedisplayskip{1pt}
\setlength\belowdisplayskip{5pt}
\mathcal{L}=f(\boldsymbol{x})+\frac{1}{2 \sigma} \sum_{i=1}^{w}\left(\left[\max \left\{\sigma g_{i}(\boldsymbol{x})+\lambda_{i}, 0\right\}\right]^{2}-\lambda_{i}{}^{2}\right)
\end{equation}
which can be solved by iterative algorithm as follows.

At iteration $\ell$, solving the unconstrained PHR augmented Lagrangian function \eqref{eq:13} yields the solution $\boldsymbol{x}^\ell$. Let $\mathcal{R}$ denote the constraint residual. If $\boldsymbol{x}^\ell$ satisfies the stop criterion (\ref{eq:14c}), the iteration is terminated, and the output $\boldsymbol{x}^\ell$ is the approximate minimum solution of the original problem. Otherwise, the Lagrange multipliers $\lambda_i^{\ell+1}$ and the penalty parameter $\sigma^{\ell+1}$ are updated according to (\ref{eq:14a}) and (\ref{eq:14b}). Here, $\varrho \in (0,1)$ is a residual reduction factor that determines whether sufficient progress in the residual is achieved, and $\eta > 1$ is the penalty growth factor.
\begin{subequations}\label{eq:14}
\setlength\abovedisplayskip{1pt}
\setlength\belowdisplayskip{5pt}
\vspace{-0em}
\begin{align}
\lambda_{i}^{\ell+1}
&=\max \!\left\{\lambda_{i}^{\ell}
+\sigma^{\ell} g_{i}\left(\boldsymbol{x}^{\ell+1}\right), 0\right\},\quad i=1, \ldots, w
\label{eq:14a}\\[4pt]
\sigma^{\ell+1}
&=
\begin{cases}
\sigma^{\ell}, & \text{if } \mathcal{R}^{\ell+1} < \varrho\mathcal{R}^{\ell},\\[2pt]\eta\,\sigma^{\ell}, & \text{otherwise}
\end{cases}
\label{eq:14b}\\[6pt]
\mathcal{R}^{\ell}
&=\left[
\sum_{i=1}^{w}
\left(
\max \left\{
-\dfrac{\lambda_{i}^{\ell}}{\sigma^{\ell}},
g_{i}\left(\boldsymbol{x}^{\ell+1}\right)
\right\}
\right)^{2}
\right]^{1 / 2}
\leq \delta
\label{eq:14c}
\end{align}
\end{subequations}

Compared to the general augmented Lagrangian multiplier method, the absence of introduced slack variables in PHR-ALM is a major advantage, making it particularly well suited for quantum algorithm reformulation to solve inequality-constrained optimization problems. 

\subsubsection{Equivalent QUBO Form and Hamiltonian}

In the QPHR-ALM framework, the new QUBO form of the objective function is obtained in the previous iteration. Therefore, the $\ell$-th iteration eigenstate $\left | \boldsymbol{x}^\ell  \right \rangle$, where $\boldsymbol{x}^\ell$ is a binary qubit string, can then be measured using a QA-QPU. As shown by Fig. \ref{QA_framework}, to obtain the quantum state solution in the next iteration, we must consider the structure of the Hamiltonian. Due to the addition of an augmented penalty term, the Hamiltonian consists of two parts, an objective Hamiltonian $\boldsymbol{H}_{O b j}$  representing an objective function $f(\boldsymbol{x})$, and the PHR-based augmented Lagrangian Hamiltonian $\boldsymbol{H}_{P H R}$  standing for the PHR augmented penalty function.

\begin{equation}\label{eq:15}
\setlength\abovedisplayskip{-5pt}
\setlength\belowdisplayskip{5pt}
\boldsymbol{H}=\boldsymbol{H}_{O b j}+\boldsymbol{H}_{P H R}=\boldsymbol{H}_{O b j}+\sum_{i=1}^{w}\boldsymbol{H}_{P H R,i}
\end{equation}

It is evident that $\boldsymbol{H}$ is not constant in each iteration. When an optimization problem is well-defined, it results in a clear expression of  $f(\boldsymbol{x})$, ensuring the invariance of $\boldsymbol{H}_{O b j}$. However, $\boldsymbol{H}_{P H R}$ is not fixed, and it represents the aggregate of the Hamiltonians corresponding to all unsatisfied constraints. Additionally, since the $\max(\cdot)$ function in (\ref{eq:13}) cannot be directly embedded in QA-QPUs, $\boldsymbol{x}^\ell$ is employed to classify the PHR augmented penalty function from (\ref{eq:12}) into two cases, determining the expression for $\boldsymbol{H}_{P H R}$
for all inequality constraints:

\begin{itemize}[leftmargin=*]
    \item \textit{Case 1}: Substituting the bit string $\boldsymbol{x}^\ell$ into $\lambda_{i}+\sigma g_{i}(\boldsymbol{x})$, if the result is less than or equal to 0, the bit string of the eigenstate $\left | \boldsymbol{x}^\ell  \right \rangle$ satisfies the $i$-th  inequality constraint. Then, according to (\ref{eq:12}), the relevant $\boldsymbol{H}_{P H R,i}$ is a diagonal matrix with component $-\left(\lambda_{i}{}^{2} / 2 \sigma\right)$. Consequently, the PHR augmented Lagrangian Hamiltonian is equivalent to a constant term, so it does not affect the optimal solution but only influences the value of objective function. Therefore, it can be neglected directly.
    \item \textit{Case 2}: When ${\lambda_{i}+\sigma g_{i}(\boldsymbol{x}^\ell)>0}$, it implies that $\left | \boldsymbol{x}^\ell  \right \rangle$ violates the current constraint, and the eigenstate result lies outside the feasible region. In this case, it is necessary to apply a related augmented penalty Hamiltonian to guide the next eigenstate sequence $\left | \boldsymbol{x}^{\ell+1}  \right \rangle$ closer to the feasible region. Consequently, QUBO-formed expression of $\boldsymbol{H}_{P H R,i}$ can be obtained by (\ref{eq:12}).
\end{itemize}

Consequently, when $\left | \boldsymbol{x}^\ell  \right \rangle$ is known, the QUBO form of the PHR augmented Lagrangian function can be derived using (\ref{eq:12}) and (\ref{eq:13}). Next, convert the QUBO form into the Hamiltonian $\boldsymbol{H}_{P H R}$ by (\ref{eq:5}) and (\ref{eq:6}). After updating the new Hamiltonian, we can use it to evolve a new quantum state $\left | \boldsymbol{x}^{\ell+1}  \right \rangle$. If the acquired quantum state satisfies the stop criterion (\ref{eq:14c}) or reaches the maximum iteration limit, the iteration can be terminated. Otherwise, proceed to adjust Lagrange multipliers and penalty coefficients relying on (\ref{eq:14a}) and (\ref{eq:14b}).

In terms of qubit efficiency, as no slack variables are introduced for inequality constraints, the qubit overhead of the QPHR-ALM algorithm is reduced  from $\underline{NT + J +\mathcal{M}\mathcal{P} + k\mathcal{F}}$ to $\underline{NT+J}$, which is independent of the increasing number of Benders' cuts during iteration and the minimum up/down time constraints. The pseudo-code is provided in Algorithm \ref{alg:alg1}.

\begin{algorithm}[H]
\caption{QPHR-ALM}\label{alg:alg1}
\begin{algorithmic}[1]
\REQUIRE{LBO-formed $f(\boldsymbol{x})$ and linear binary constraints}
\ENSURE{Optimal quantum state $\left|\boldsymbol{x}^{*}\right\rangle$}
\STATE $ \textbf{Initialization: } \!\ell \!\gets\!\! 1, \sigma_{0}\!>\!0,\lambda_{i}^{0} \!\gets\! 0, \eta\!>\!1,\varrho\!\in\! (0,1), \epsilon>0  \ \forall i$
\WHILE {(\ref{eq:14c}) does not satisfy $\textbf{or}$ $\ell<\ell^{max }$}
\STATE Obtain  $\left|\boldsymbol{x}^{\ell}\right\rangle$  using QA algorithms
\FOR {each inequality constraint}
        \IF {$\lambda_{i}+\sigma g_{i}\left(\boldsymbol{x}^{\ell}\right)>0$}
        \STATE Add an augmented penalty Hamiltonian using the QUBO form based on (\ref{eq:5}), (\ref{eq:6}) and (\ref{eq:12})
        \ELSE
        \STATE Keep Hamiltonian unchanged
        \ENDIF
\ENDFOR
\STATE Renew the QUBO form of the objective function and the corresponding Hamiltonian 
\STATE Update coefficients using (\ref{eq:14a}) and (\ref{eq:14b})
\STATE $\ell \gets \ell+1$
\ENDWHILE
\end{algorithmic}
\label{alg1}
\end{algorithm}

\subsection{Quantum-based PHR-ADMM Algorithm}
\label{section:III C}
Although the QPHR-ALM method eliminates slack variables and reduce the qubit overhead to
$NT+J$, it remains inadequate for practical SUC instances. For example, when scheduling a 24-hour operation on IEEE bus-118 system with 54 generator units hourly, the master problem, after applying BD, would involve at least 1296 integer variables. Since it is impractical to rely on a single quantum computer for the entire computational workload, the ADMM algorithm is adopted for solving SUC master problem. This approach decomposes a large-scale LBO-formatted integer programming master problem into several smaller sub-problems that can be efficiently managed by smaller scale QPUs, and the solution to the large global problem is obtained by coordinating the solutions to all the sub-problems. 

In particular, for multi-unit, multi-period SUC, the D-ADMM approach decomposes the master problem either by \emph{generator units} or \emph{time horizon}. Under unit-wise decomposition, the states for all time periods of each unit are grouped into the same block, followed by a block accouting for the binary encoded continuous variable $\Upsilon^k$ with $J$ binary variables. As $T$ is conventionally fixed as 24 in power system operation, this setting aligns with the worst possible qubits number when $N\gg T$. Moreover, the ADMM block size must respect hardware embedding limits to avoid large minor embeddings that can increase chain breaks; for the current D-Wave Advantage QPU, each subproblem is preferably restricted to roughly 30–32 binary variables \cite{willsch2022benchmarking}. From this perspective, decomposing by generator units is the more reasonable and hardware-aware choice.

The multi-block D-ADMM algorithm for the compact master problem \eqref{eq:8} can be expressed as
\begin{equation}\label{eq:16}
\setlength\abovedisplayskip{1pt}
\setlength\belowdisplayskip{5pt}
\begin{aligned}
\min & \sum_{m=1}^{M} f_{m}\left(\boldsymbol{x}_{m}\right) \\
\text { s.t. } & \sum_{m=1}^{M} g_{i, m}\left(\boldsymbol{x}_{m}\right) \leq 0 \text { for } i=1, \ldots, w
\end{aligned}
\end{equation}
where $M$ is the number of ADMM blocks. Based on a similar derivation process in Section \ref{section:III B}, we can easily obtain the PHR-based augmented Lagrangian function for the D-ADMM form after eliminating the slack variables as, 
\begin{multline}\label{eq:17}
\setlength\abovedisplayskip{1pt}
\setlength\belowdisplayskip{5pt}
\!\!\!\!\!\!\mathcal{L}_{D}\!=  \frac{1}{2 \sigma} \!\sum_{i=1}^{w}\left(\!\left[\max \left\{\sigma \!\sum_{m=1}^{M} g_{i, m}\left(\boldsymbol{x}_{m}\right)\!+\lambda_{i}, 0\right\}\right]^{2}\!\!-\!\lambda_{i}{}^{2}\!\right)\\
 +\sum_{m=1}^{M} f_{m}\left(\boldsymbol{x}_{m}\right)
\end{multline}

Next, the integer variables within each block can be updated step-by-steps in (\ref{eq:18a}) using a QA-QPU. After updating all state variables, the Lagrange multipliers and penalty coefficients are adjusted according to (\ref{eq:18b}) and (\ref{eq:18c}), respectively. The error is evaluated to determine whether it satisfies the convergence criteria in (\ref{eq:18d}).
\begin{subequations}\label{eq:18}
\setlength\abovedisplayskip{1pt}
\setlength\belowdisplayskip{5pt}
\begin{align}
\boldsymbol{x}_{m}^{\ell+1}=&\arg \min _{\boldsymbol{x}_{m}^{\ell}} \mathcal{L}_{D}(\boldsymbol{x}_{1}^{\ell+1}, \ldots, \boldsymbol{x}_{m-1}^{\ell+1}, \boldsymbol{x}_{m}^{\ell}, \boldsymbol{x}_{m+1}^{\ell}, \ldots, \boldsymbol{x}_{M}^{\ell}, \nonumber\\
&\lambda^{\ell}, \sigma^{\ell} ) \text { for } m=1, \ldots, M \label{eq:18a}\\
\lambda_{i}^{\ell+1}=&\max \left\{\lambda_{i}^{\ell}+\sigma^{\ell} \sum_{m=1}^{M} g_{i, m}\left(\boldsymbol{x}_{m}^{\ell+1}\right), 0\right\}\! \text { for } i=1, \ldots, w \label{eq:18b}\\
\sigma^{\ell+1}=&
\begin{cases}
\sigma^{\ell}, & \text{if } \mathcal{R}^{\ell+1} < \varrho\mathcal{R}^{\ell},\\[2pt]
\eta\,\sigma^{\ell}, & \text{otherwise}
\end{cases} \label{eq:18c}\\
\mathcal{R}^{\ell}\!=\!&\left[\sum_{i=1}^{w}\!\!\left(\max \!\left\{\!-\frac{\lambda_{i}^{\ell}}{\sigma^{\ell}}, \!\sum_{m=1}^{M}\! g_{i, m}\!\left(\boldsymbol{x}_{m}^{\ell+1}\right)\!\right\}\!\right)^{\!2}\right]^{\!1 / 2} \!\!\leq\! \delta \label{eq:18d}
\end{align}
\end{subequations} 

Notably, as each block in \eqref{eq:18a} can be serially solved on QPUs, the total QPU qubit overhead further reduces from $NT+J$ to $\underline{\max\{T,J\}}$ as a constant.

\subsection{Summary on the BD and QPHR-ADMM Framework}\label{section:III D}

\begin{figure}[!t]
\centering
\includegraphics[width=3.3 in]{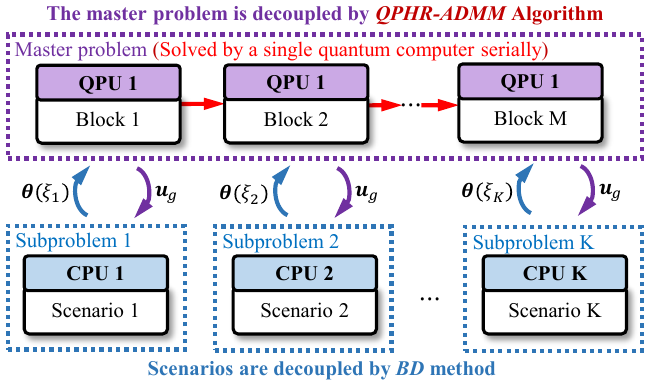}
\caption{The hybrid quantum-classical algorithm for SUC (The purple region is decoupled using the proposed QPHR-ADMM algorithm).}
\label{fig_2}
\vspace{-1em}
\end{figure}

Our proposed QPHR-ADMM algorithm, powered by PHR Augmented Lagrangian and quantum ADMM, is summarized in Fig. \ref{fig_2} and the overall saving on qubits numbers is summarized in Table \ref{tab:summary}, which is \emph{independent} of the number of generators and the number of BD iterations. The algorithm begins by decoupling the SUC problem using BD. This decoupling serves two primary purposes for both CPU and QPU efficiency. First, it ensures that all scenarios within the second stage are independent and continuous in the sub-problems, which can be efficiently solved by allocating dedicated CPUs to each scenario \emph{in parallel}. Second, the first-stage master problem is decomposed by ADMM, allowing for sequential processing of each block on a single quantum processor. Each block is then solved by the PHR Augmented Lagrangian method without occupying extra qubits for slack variables introduced by Benders' cuts. The binary unit decisions from QPHR-ADMM iterations are then integrated into subproblems to optimize power outputs and update dual variables. The calculation of subproblems contributes to tightening the UB of BD. The process repeats until the BD bounds converge within a small tolerance.

\begin{table}[h]
    \centering
    \renewcommand{\arraystretch}{1.1}
    \vspace{-0.5em}
    \caption{Summary on the qubit number of QPHR-ADMM under BD framework.}
    \begin{tabular}{cc}
        \toprule[1.2pt]
        \hline
        \textbf{Method} & \textbf{\makecell{Qubit Number}} \\\hline
        Basic QA & $NT + J +\mathcal{M}\mathcal{P} + k\mathcal{F}$\\
        QPHR-ALM & $NT+J$\\
        QPHR-ADMM & $\max\{T, J\}$\\
        \bottomrule[1pt]
    \end{tabular}
    \label{tab:summary}
\end{table}

\subsection{Discussion on the Algorithm Convergence}\label{section:III E}

The convergence of the proposed algorithm depends on convergence properties of its individual components, inclduing BD, QA solver, augmented Lagrangian, and D-ADMM. First, BD guarantees finite termination and global optimality for mixed-integer programs under standard assumptions, notably exact subproblem solutions and the generation of valid cuts \cite{geoffrion1972generalized}. For the QA solver on the master problem, convergence guarantees are primarily available in idealized adiabatic regimes. When the annealing process is sufficiently slow and appropriate regularity conditions hold, the final state remains close to the ground state, with potentially exponential error decay under stronger smoothness assumptions\cite{Albash2018AdiabaticQC}. In our setting, the Benders master problem in this work is discrete and non-strongly convex; recent results \cite{wang2024customized} show that, for block-structured integer programs, an augmented Lagrangian formulation with a sufficiently large but finite penalty parameter can yield an exact relaxation and valid lower bounds, provided that the subproblems are solved accurately. At last, for convex problem, the convergence of ADMM typically requires the penalty $\sigma$ to be chosen within a suitable range and the step size factor $\eta$ to be sufficiently small  \cite{Hong2017linearCADMM, Han2012noteADMM}. By contract, ADMM lacks general convergence guarantees in nonconvex or discrete settings and is therefore used here as a coordination mechanism rather than as a provably convergent solver. Nevertheless, empirical evidence in power system applications suggests that stable performance can be achieved by initializing $\sigma$ conservatively and increasing it gradually \cite{wuijts2023new, mhanna2018adaptive}.

\section{Simulations} \label{section:IV}

This section validates the reliability of the proposed algorithm through synthetic binary problems as well as SUCs based on a 4-generator power system demonstrator and the IEEE bus-118 system. The proposed and baseline QC algorithms are tested on the QPU solver, D-Wave Advantage\_system 4.1. We also use the Gurobi solver on a classical computer with an Intel Core i7-12700H processor as reference. To start, we verify the convergence of the proposed QPHR-ADMM on a synthetic MIP instance in Appendix~\ref{section:IV A}, which also demonstrates that QA yields more accurate solutions than QAOA in this setting.

\subsection{4 Generators Scenario-based SUC Example}

The power system considered in this section consists of four generators, an equivalent wind generator, and a load. The output of the wind turbine is influenced by wind speed over a 24-hour period, which follows a Weibull probability distribution. Additionally, the load demand is uncertain and is assumed to follow a Beta distribution\cite{Maneesh2020AnOptimalMVCNL}.

\subsubsection{Theoretical Analysis on Qubit Overhead}

Considering that the time horizon of the SUC is $24$ hours, the master problem after BD involves $24\times4=96$ decision variables for the total commitment statuses. Additionally, $J=12$ and $\mathcal{F}=13$ are set to allocate qubits for discretizing the LB and slack variables, with the corresponding precision coefficients specified as $0.004$ and $0.005$, respectively. Therefore, If the minimum up/down time constraints are temporarily ignored, a total of $108 + 13k$ qubits are required in the basic QA where $k$ is the number of Bender's iterations. In contrast, since no binary-encoded slack variables are introduced, the number of qubits required for QPHR-ALM can be determined as $108$. In contrast, based on the QPHR-ADMM algorithm, we decompose the master problem into five unit-based blocks. The first four blocks contain the on$/$off decision variables for each generator over the $24$-hour period, while the last block is used to compute the discretized LB value. As a result, these five blocks require $24$, $24$, $24$, $24$, and $12$ qubits. As the D-ADMM framework can be implemented in serial, it implies that the QPHR-ADMM algorithm requires only a QPU with a capacity of $24$ qubits to finish tasks that would otherwise demand $108 + 13k$ qubits.  

If the minimum up- and down-time requirements are set to \( T_g^{U} = [6, 4, 3, 2] \) and \( T_g^{D} = [6, 4, 3, 2] \) for 4 units, respectively, 
the four generators will give rise to a total of 162 inequality constraints.
If the same slack-variable setting that $\mathcal{P}=13$ is applied to each constraint, the resulting binary slack variables would require 2106 qubits in total, which is far beyond the capability of the Basic QA. In contrast, QPHR-ADMM still requires only 24 qubits.

\subsubsection{Convergence and Sensitivity Analysis}

\begin{figure}[h]
\centering
\includegraphics[width=3.5 in]{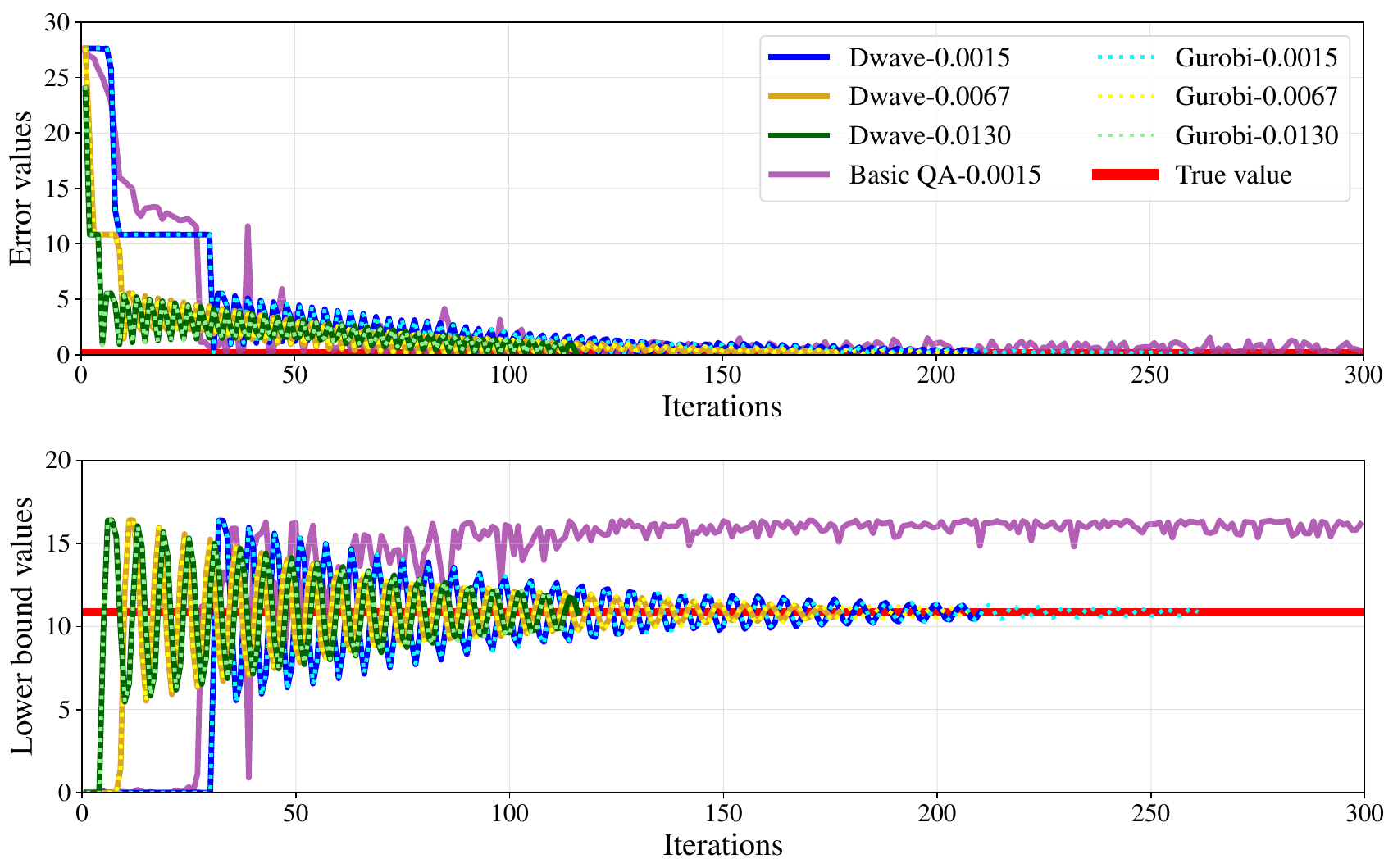}
\vspace{-1.7em}
\caption{Stopping criteria error (top) and BD lower bound (bottom) versus ADMM iterations within the first BD iteration for \( \sigma^{0}\in\{0.0015, 0.0067, 0.0130\} \). Solid lines: D-Wave; dotted lines: Gurobi; purple: Basic QA (\( \sigma^{0}=0.0015 \)); red line: true value. }
\label{fig_5}
\vspace{-0.5em}
\end{figure}

Fig. \ref{fig_5} shows that the trends in the LB values and the corresponding errors obtained using the QPHR-ADMM method (with different parameter $\sigma^0$) and Basic QA during the first iteration of BD for the $100$ sampled scenarios. It should be emphasized that the results in Fig. \ref{fig_5} are obtained without minimum up- and down-time constraints. Including these constraints increases the qubit requirement beyond the minor-embedding capability of current quantum annealing hardware, making Basic QA infeasible.

\begin{figure}[t]
\centering
\includegraphics[width=3 in]{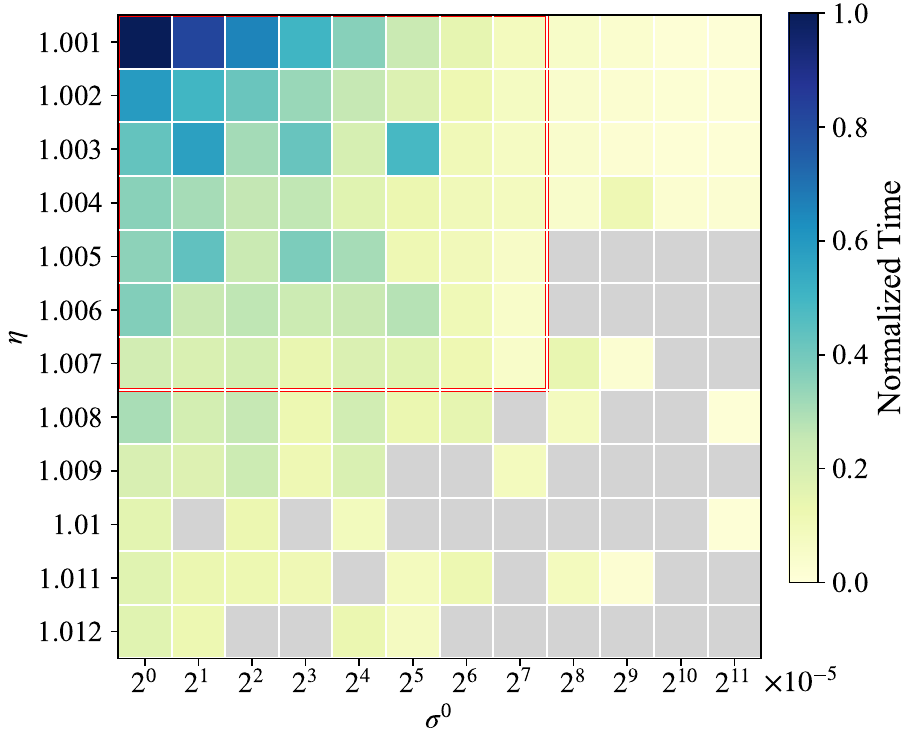}
\vspace{-0em}
\caption{The stability map for 4 generator UC problem with 100 scenarios (Gray blocks mean non-convergence)}.
\label{fig_sensitivity}
\vspace{0em}
\end{figure}

QPHR-ADMM demonstrates satisfactory convergence performance. For example, as shown in Fig. \ref{fig_5}, the LB value gradually narrows down to the optimal value, and the stopping criteria error eventually approaches zero during the QPHR-ADMM iterations. It indicates that the master problem will converge to the correct solution. Additionally, This figure also demonstrates the impact of the initial penalty parameters on convergence performance. For instance, a larger $\sigma^0$ can reduce iterations, thereby saving computational time. However, due to the uncertain convergence of QPHR-ADMM, setting a smaller $\sigma^0$ is necessary to ensure more reliable convergence, as discussed before. Moreover, the measured results from basic QA often fail to converge to the correct ground-state solution, leading to erroneous LB values as shown by the purple curves. Such issues occur due to the high qubit demand, which leads to longer embedding chains. The increased chain length reduces the effective chain strength of certain qubits, resulting in broken chains in current D-Wave QA-QPUs \cite{Elijah2023ComparingTG}. Consequently, employing QPHR-ADMM not only lowers qubit overhead per annealing process but also plays a critical role in maintaining solution accuracy, even under limited qubit resources.

Table \ref{tab3} illustrates the convergence of upper and lower bounds in the BD method under different numbers of scenarios, sets of constraints, and wind penetration level. The UB results are obtained by aggregating all the subproblems, which are processed exclusively by classical computers. The LB can be computed using either QPUs or CPUs for reference. As shown in the table, after introducing time-coupled start-up/down constraints and higher wind power penetration, the proposed method is still able to converge. Note that there exists a slight discrepancy between the LBs computed by Gurobi and the proposed QPHR-ADMM, due to the precision level of the discretized LB being set at $0.004$. 

At last, Fig. \ref{fig_sensitivity} presents the sensitivity of convergence time on parameters $\sigma^0$ and $\eta$ in QPHR-ADMM with minimum up/down constraints. As outlined in Section \ref{section:III E}, smaller penalty coefficients $\sigma^0$ and step size factors $\eta$ tend to promote more stable convergence, whereas increasing them can substantially reduce run time and accelerate convergence; however, excessively large values may result in non-convergence. Therefore, we recommend selecting the parameters within the red box.

\subsubsection{Timing Breakdown of the Hybrid Quantum-Classical Workflow}

To clarify the practical runtime of the proposed hybrid workflow, Table \ref{tab} reports the BD-level timing breakdown of the 4-unit case. Here, \(T_{\mathrm{QPU}}\) is the QPU access time reported by the D-Wave Solver API, and \(T_{\mathrm{orch}}\) is the orchestration time outside direct QPU access. The end-to-end (E2E) time of the quantum workflow, \(T_{\mathrm{E2E}}\), is given by the sum of these two terms. Other timing components are defined in Appendix \ref{app}.

\begin{table*}[t]
\centering
\caption{Timing breakdown of the hybrid workflow across BD iterations. All values are in seconds.}
\label{tab}

\setlength{\tabcolsep}{3pt}
\renewcommand{\arraystretch}{1.05}{
\begin{tabular}{cccccccccccc}

\toprule[1.1pt]
\hline
\multicolumn{2}{c}{\textbf{Iterations}} &
\multicolumn{9}{c}{\textbf{Quantum workflow (Master Problem)}} &
\multicolumn{1}{c}{\textbf{Classical computing (Subproblems)}} \\
\cmidrule(lr){1-2}
\cmidrule(lr){3-11}
\cmidrule(lr){12-12}
\shortstack{\textbf{BD}\\\textbf{iter.}} &
\shortstack{\textbf{Total ADMM}\\\textbf{iter.}} &
\(T_{\mathrm{QPU}}\) &
\(T_{\mathrm{prog}}\) &
\(T_{\mathrm{samp}}\) &
\(T_{\mathrm{anneal}}\) &
\(T_{\mathrm{readout}}\) &
\(T_{\mathrm{delay}}\) &
\(T_{\mathrm{post}}\) &
\(T_{\mathrm{orch}}\) &
\(T_{\mathrm{E2E}}\) &
\shortstack{\textbf{CPU parallel time}} \\
\midrule
0 & --  & --    & --    & --    & --   & --   & --   & --   & --     & --     & 0.0423 \\
1 & 129 & 22.94 & 10.17 & 12.77 & 1.61 & 9.48 & 1.67 & 1.49 & 370.53 & 393.46 & 0.0923 \\
2 & 90  & 15.86 & 7.09  & 8.77  & 1.13 & 6.47 & 1.18 & 1.20 & 262.90 & 278.76 & 0.0659 \\
\bottomrule
\end{tabular}
}
\end{table*}

Table \ref{tab} shows that \(T_{\mathrm{orch}}\) is much larger than \(T_{\mathrm{QPU}}\). For example, in BD iteration 1, \(T_{\mathrm{QPU}}\) is 22.94 s, and \(T_{\mathrm{orch}}\) is 370.53 s. The annealing time is also small, with \(T_{\mathrm{anneal}}=1.61\) s in the same iteration. Therefore, the current E2E runtime is mainly dominated by orchestration latency rather than direct QPU execution.

\begin{figure}[!t]
\centering
\includegraphics[width=3.5 in]{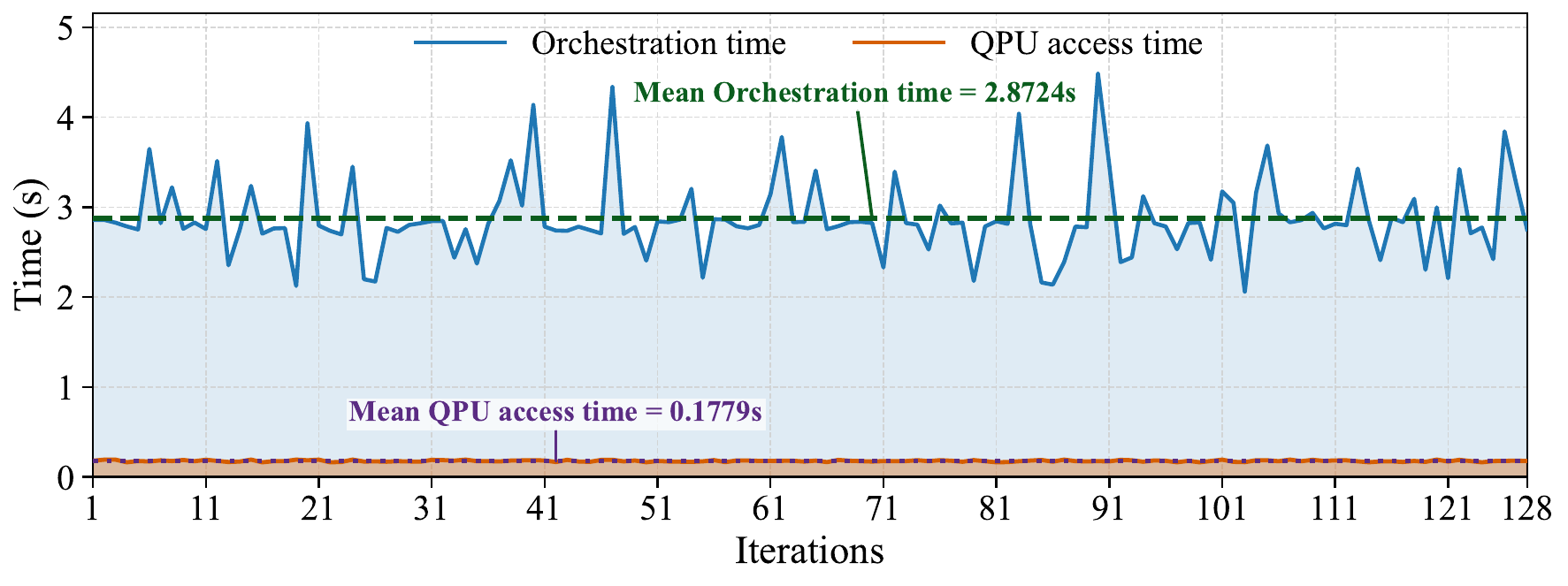}
\caption{Time variation of orchestration and QPU access in one BD iteration over ADMM iterations.}
\label{fig_BD_Time}
\vspace{-1em}
\end{figure}

Fig. \ref{fig_BD_Time} further shows the per-ADMM-iteration timing within one BD iteration. The average QPU access time is 0.178 s, and the average orchestration time is 2.872 s. This is consistent with the BD-level results in Table \ref{tab} and shows that the timing gap appears throughout the ADMM iterations. Hence, the larger E2E time mainly reflects the external overhead outside direct QPU access and cannot directly represent the execution efficiency of the quantum module. The detailed reasons for this gap are discussed in Section \ref{sec:quantum_advantage}.

\begin{table}[t]
\centering
\caption{The BD Results for SUC Problems with 4 Generators.}
\label{tab3}
\setlength{\tabcolsep}{3pt}
\renewcommand{\arraystretch}{1.3}

{
\begin{tabular}{cccccccc}
\hline
\hline
\multicolumn{8}{l}{\textbf{Assumptions: No minimum up/down time constraints}}  \\
\hline
\hline
 & \multicolumn{3}{c}{$10$ scenarios} &  & \multicolumn{3}{c}{$100$ scenarios} \\ \cline{2-4} \cline{6-8}
\shortstack{Iteration\\No.} &
\shortstack{UB\\(Gurobi)} &
\shortstack{LB\\(Gurobi)} &
\shortstack{LB\\(Dwave)} &
&
\shortstack{UB\\(Gurobi)} &
\shortstack{LB\\(Gurobi)} &
\shortstack{LB\\(Dwave)} \\
\hline
$0$& $27.856$ & $0$ & $0$ & & $27.644$ & $0$ & $0$ \\
$1$& $11.501$ & $11.056$ & $11.056$ & & $11.290$ & $10.844$ & $10.844$ \\
$2$& $11.501$ & $11.501$ & $11.500$ & & $11.290$ & $11.290$ & $11.288$ \\
\hline
\hline
\multicolumn{8}{l}{\textbf{Assumptions: With} $T_g^{U}=[6,4,3,2]$ and $T_g^{D}=[6,4,3,2]$} \\
\hline
\hline

 & \multicolumn{3}{c}{$10$ scenarios} &  & \multicolumn{3}{c}{$100$ scenarios} \\ \cline{2-4} \cline{6-8}
\shortstack{Iteration\\No.} &
\shortstack{UB\\(Gurobi)} &
\shortstack{LB\\(Gurobi)} &
\shortstack{LB\\(Dwave)} &
&
\shortstack{UB\\(Gurobi)} &
\shortstack{LB\\(Gurobi)} &
\shortstack{LB\\(Dwave)} \\
\hline
$0$& $27.856$ & $0$ & $0$ & & $27.644$ & $0$ & $0$ \\
$1$& $11.896$ & $11.367$ & $11.368$ & & $11.665$ & $11.136$ & $11.136$ \\
$2$& $12.201$ & $11.957$ & $11.956$ & & $11.823$ & $11.698$ & $11.700$ \\
$3$& $12.058$ & $11.924$ & $11.924$ & & $11.665$ & $11.665$ & $11.664$ \\
$4$& $12.040$ & $11.929$ & $11.928$ & & --- & --- & --- \\
$5$& $11.896$ & $11.896$ & $11.896$ & & --- & --- & --- \\

\hline
\hline
\multicolumn{8}{l}{\textbf{Assumptions: With} $T_g^{U}=[6,4,3,2]$, $T_g^{D}=[6,4,3,2]$ and $1.1 P^{W i n d}$} \\
\hline
\hline
 & \multicolumn{3}{c}{$10$ scenarios} &  & \multicolumn{3}{c}{$100$ scenarios} \\ \cline{2-4} \cline{6-8}
\shortstack{Iteration\\No.} &
\shortstack{UB\\(Gurobi)} &
\shortstack{LB\\(Gurobi)} &
\shortstack{LB\\(Dwave)} &
&
\shortstack{UB\\(Gurobi)} &
\shortstack{LB\\(Gurobi)} &
\shortstack{LB\\(Dwave)} \\
\hline
$0$& $27.660$ & $0$ & $0$ & & $27.442$ & $0$ & $0$ \\
$1$& $11.761$ & $11.170$ & $11.172$ & & $11.532$ & $10.933$ & $10.932$ \\
$2$& $12.054$ & $11.822$ & $11.824$ & & $11.681$ & $11.562$ & $11.560$ \\
$3$& $11.915$ & $11.790$ & $11.788$ & & $11.532$ & $11.532$ & $11.532$ \\
$4$& $11.901$ & $11.794$ & $11.796$ & & --- & --- & --- \\
$5$& $11.761$ & $11.761$ & $11.760$ & & --- & --- & --- \\
\hline
\end{tabular}
} 

\vspace{-1.5em}
\end{table}

\begin{figure}[t]
\centering
\includegraphics[width=3.5 in]{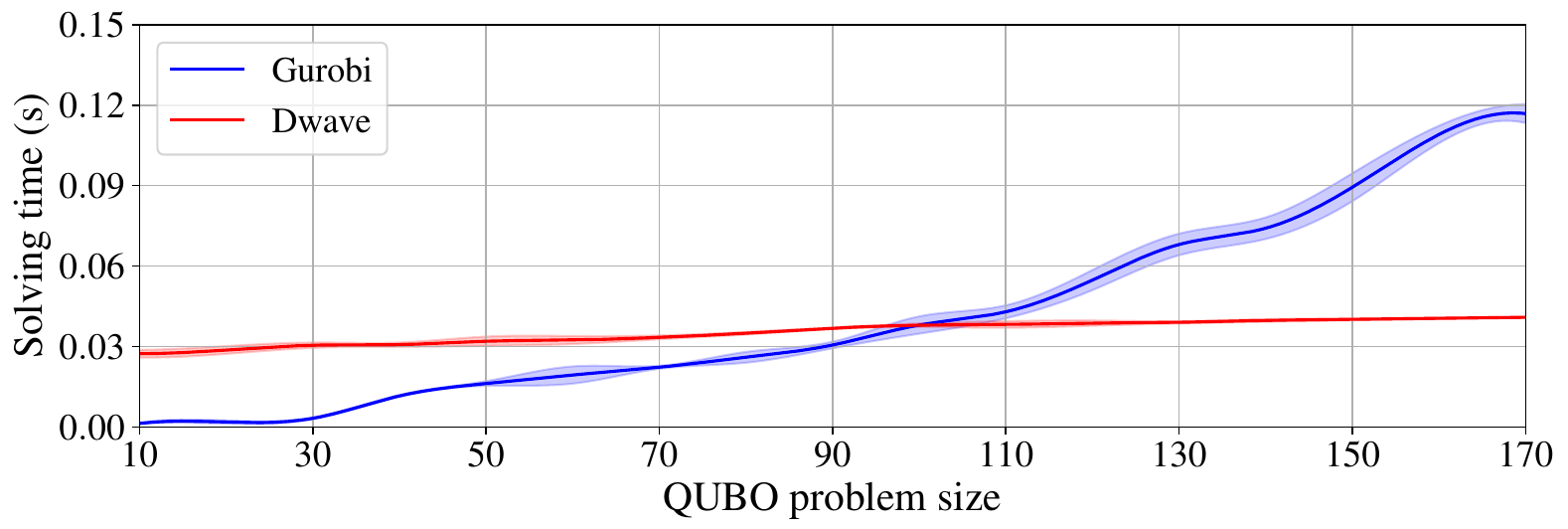}
\vspace{-1.7em}
\caption{Average direct QPU access time cost for different QUBO sizes (equivalent to the qubit requirement of a single ADMM block).}
\label{fig_6}
\vspace{-1em}
\end{figure}

\subsection{Large-scale SUC Problem}

This section evaluates the performance of QPHR-ADMM on large-scale SUC problems using IEEE bus-118 system with 54 generating units. For modeling simplicity, the minimum up/down time of all generating units are assumed to be 4 hours. Table \ref{tab_54gens} summarizes the optimization results of the SUC problem using classical and quantum computing methods. The problem contains 1296 binary decision variables, and the basic QA is infeasible due to hardware limitations. However, the QPHR-ADMM enables solving the problem with only 24 qubits. In terms of solution quality, the classical BD method using Gurobi achieves a total cost of 1728.88~k\$. The proposed QPHR-ADMM attains an average cost of 1734.26~k\$, with a standard deviation of 1.12~k\$ over multiple runs. This corresponds to a cost increase of approximately 0.31\% compared with the classical BD solution. The observed gap mainly results from occasional QA errors in unit commitment decisions, which lead to suboptimal on/off statuses in a small number of time periods. These results demonstrate that QPHR-ADMM maintains convergence behavior over large system and delivers near-optimal solutions while significantly reducing the quantum resource requirement.

\begin{table}[t]
\setlength\tabcolsep{2.2pt}
\begin{center}
\caption{Cost Comparison Between Classical and Quantum Method.}
\label{tab_54gens}
\renewcommand{\arraystretch}{1.1}
\begin{tabular}{ccc|cc}
\toprule[1.2pt]
\hline
\multicolumn{3}{c|}{ \textbf{Classcial Computing}} & \multicolumn{2}{c}{ \textbf{Quantum Computing}} \\ \hline
\makecell{Cost \\(Gurobi)} 
& \makecell{Cost \\(Gurobi + BD)}& \makecell{Binary\\Variables} 
& \makecell{Average Cost \( \pm \) std \\(QPHR-ADMM + BD)}& \makecell{Qubit\\Overhead}  \\ \hline
\(1728.80\,k\$ \)
& \(1728.88\,k\$\)&1296 
& \(1734.26 \pm 1.12\,k\$\)&24 \\
\bottomrule[1pt]
\end{tabular}
\end{center}
\vspace{-1.5em}
\end{table}

\subsection{Discussion and Prospects for Quantum Advantage}
\label{sec:quantum_advantage}

In the context of QC, its primary potential advantage lies in accelerating computations for power system scheduling. This section discusses such prospects under both current hardware constraints and anticipated future developments. Fig. \ref{fig_6} illustrates the solving time of D-Wave QA processors for QUBO problems of varying sizes, considering only direct QPU access time and ignoring solution errors in larger-scale annealing. 

For the smaller-scale QUBO problems, classical computers exhibit a clear advantage in computational speed over quantum processors. However, as the problem size increases, the computational time for classical methods escalates significantly due to the NP-hard nature of integer programming. In contrast, the measured QPU access time of QA shows relatively weak sensitivity to problem size growth. These empirical observations suggest that, beyond a certain problem scale (e.g., on the order of $100$ equivalent qubits), quantum annealing may become competitive with classical approaches in terms of solver execution time, provided that solution quality is maintained.

However, QPU access time reported in Fig. \ref{fig_6} does not represent the E2E computational cost of QA. This is further supported by the timing breakdown in Fig. \ref{fig_BD_Time}, where direct QPU execution accounts for only a small portion of the current QA workflow. The much larger orchestration time mainly comes from minor-embedding construction, remote communication with the QPU service, task scheduling, and queuing. In contrast, classical solvers such as Gurobi can be deployed locally with negligible communication latency. Therefore, under current hardware and access conditions, QA-based computation does not yet show a clear E2E timing advantage over state-of-the-art classical solvers.

In addition to the timing overhead, current hardware limitations also constrain the implementable problem size and solution stability. For large-scale SUC instances, the QUBO graph is usually too large and too dense to be directly mapped onto the sparse hardware graph of current D-Wave QPUs. Hence, minor embedding is required, which may introduce long embedding chains and broken-chain risks. QPHR-ADMM mitigates this issue from the algorithmic side. Specifically, it eliminates the qubit overhead caused by binary-encoded slack variables and decomposes the full master problem into smaller unit-level QUBO blocks. This reduces the graph size that must be embedded in each QPU call. As a result, minor embedding becomes easier, and solution stability on current hardware can be improved.

Building on this algorithm, future embedding methods and hardware improvements can further strengthen the scalability and stability of the proposed framework. For the embedding-related overhead, an all-to-all embedding strategy \cite{pelofske2025comparing} can be used to construct a reusable embedding for a larger QUBO interaction graph that covers the possible QUBO structures generated during BD and ADMM iterations. In this case, the embedding does not need to be reconstructed for each QUBO instance. Instead, only the corresponding QUBO coefficients need to be updated. Improved QPU connectivity can further shorten embedding chains, reduce minor-embedding difficulty, and lower the broken-chain rate. In the ideal case of fully connected logical connectivity, QUBO graphs could be embedded more directly and efficiently. For the access-related overhead, local deployment of QA devices, lower-latency access interfaces, and tighter CPU--QPU integration may reduce communication, scheduling, and queuing costs.

Overall, the proposed QPHR-ADMM framework is meaningful under current and future hardware constraints because it reduces the qubit overhead. As QPU connectivity improves, broken-chain rates decrease, and the number of reliable logical qubits increases, the framework is expected to support larger block sizes while maintaining solution stability. Together with embedding reuse and tighter CPU--QPU integration, these advances can narrow the gap between QPU access time and E2E wall-clock time, making the proposed framework more practical for future QA-based large-scale power system optimization.

\section{Conclusion}\label{section:V}

This paper investigates the feasibility of solving scenario-based SUC through quantum computation. Building on the classic BD framework, we propose a novel QPHR-ADMM algorithm that integrates the principles of QPHR-ALM and D-ADMM. The proposed approach enables independent acceleration of the master problem on a quantum annealer, while the subproblems are solved in parallel on CPUs. Additionally, the algorithm  mitigates potential quantum solver failures and reduces qubit overhead, maintaining it at a constant level rather than scaling linearly with the number of Benders iterations, generators, and additional constraints. By demonstrating the applicability of QA to address synthetic and SUC problems, this study establishes a foundational framework for addressing complex mixed-integer inequality constrained optimizations in power system operation. Furthermore, the framework can be extended to other BD-based power system optimization problems with large-scale binary master variables, such as robust UC or transmission expansion planning, although problem-specific modeling and convergence challenges may require further investigation.

\appendix

\subsection{Binary Integer Programming Example}
\label{section:IV A}

QPHR-ADMM obviates the need for continuous slack variables when handling integer inequality constraints, thereby enabling the optimization problem to achieve optimal solutions entirely within the QPU, without recourse to classical computation. The convergence of the algorithm is first validated by the following example. 
\begin{subequations}\label{eq:19}
\setlength\abovedisplayskip{1pt}
\setlength\belowdisplayskip{5pt}
\begin{align}
\min_{\boldsymbol{x}\in\{0,1\}^6} &6 x_{1}+3 x_{2}-5 x_{3}-6 x_{4}+4 x_{5}-7 x_{6} \label{eq:19a}\\
\text { s.t. }&-2 x_{2}-2 x_{5}-x_{6}+3 \leq 0 \label{eq:19b}\\
&-x_{1}+x_{3}-x_{4}+2 x_{6} \leq 0 \label{eq:19c}\\
&-x_{1}+x_{3}+x_{4} \leq 0 \label{eq:19d}
\end{align}
\end{subequations}

In this case study, the expression of the objective function follows the structure of the master problem discussed in Section \ref{section:III B}, specifically focusing on integer programming without quadratic terms. Furthermore, all constraints are linear inequality constraints with binary variables and the qubit string can be denoted as $\left | x_1 x_2 x_3 x_4 x_5 x_6  \right \rangle$. To test the effectiveness of QPHR-ADMM, three variants of \eqref{eq:19} are considered: \textit{unconstrained}, \textit{single inequality constraint}, and \textit{multiple inequality constraints} optimization, which can mimic the progressive introduction of Benders cuts when solving the SUC problem. 

Theoretically, basic QA requires 6 qubits for each binary variables, and each inequality constraint will introduce one slack variable that needs to be encoded by $\mathcal{F}$ binary variables. Refering to Table \ref{tab:summary}, the total qubits occupation is $6+k\mathcal{F}$ where $k$ is the number of inequality constraints in this case study and $J$ is omitted as there is no continuous variable. For the proposed QPHR-ADMM, the default parameters are set as $\eta=1.05$ and $\delta=0.01$, and the entire problem is divided into three ADMM blocks, each containing $2$ qubits. These blocks are updated sequentially, and upon completion of this process, the coefficients will be updated.

The test results of QPHR-ADMM are listed in Table \ref{tab1}. It demonstrates that QPHR-ADMM can achieve highly precise solutions for all constraints settings. Notably, the ADMM algorithm can reduce the constant qubit occupation from $6$ to $2$, and unlike the basic QA, the qubit overhead will not increase as the number of constraints increases. This capability underscores its potential to effectively address the master problem of SUC.

\begin{figure}[!t]
\centering
\includegraphics[width=3.5in]{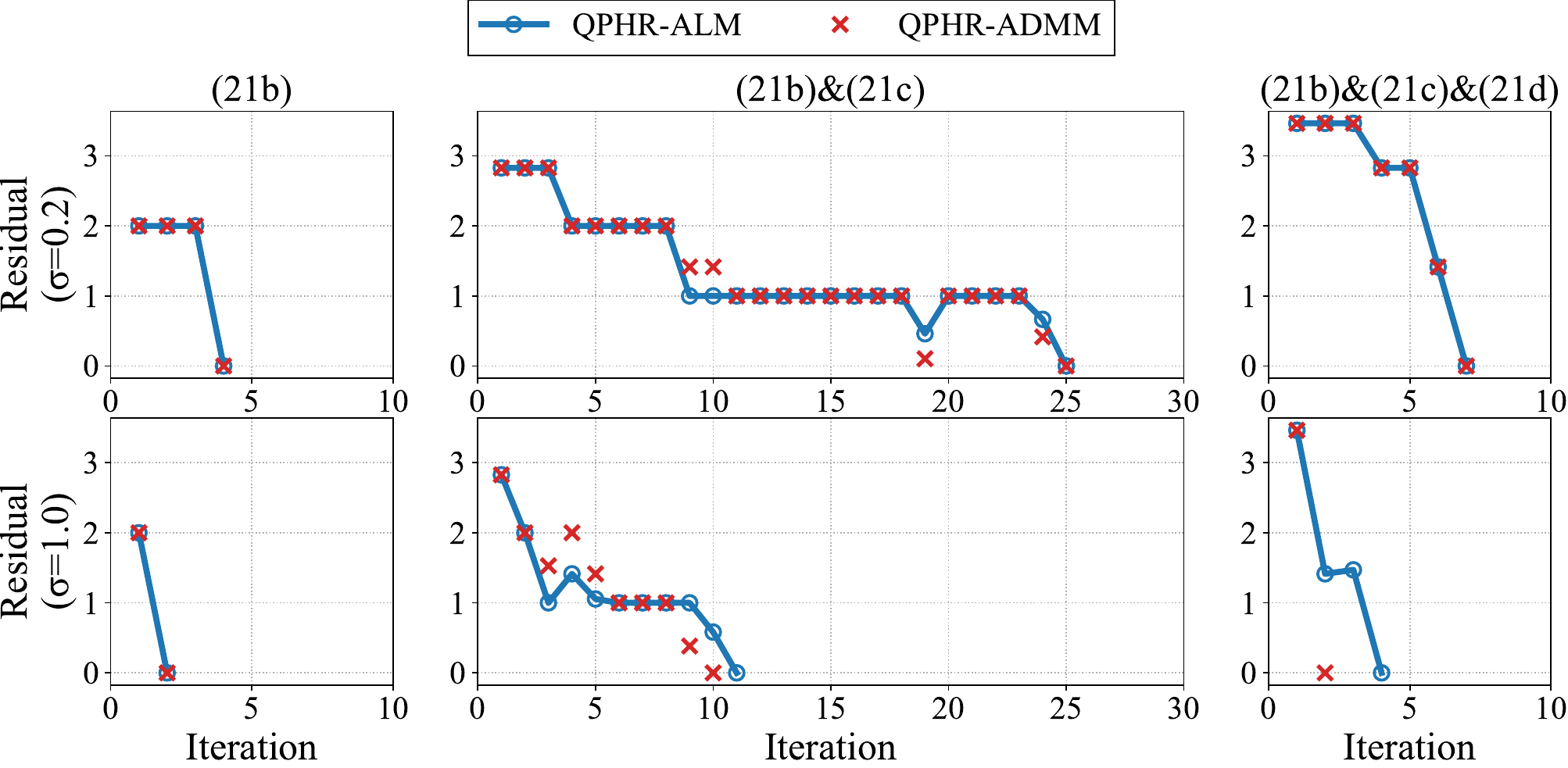}
\vspace{-1.5em}
\caption{Residual Comparison of QPHR-ALM and QPHR-ADMM.}
\label{fig_QPHR_ALM_ADMM}
\vspace{-1em}
\end{figure}

In addition, to examine whether the ADMM-based decomposition affects computational accuracy, we compare the iterative residuals of QPHR-ADMM and QPHR-ALM (without ADMM), as shown in Fig. \ref{fig_QPHR_ALM_ADMM} under two penalty coefficient $\sigma$s. The results indicate that, for all considered cases and parameter configurations, the residuals of both methods converge to zero. As a result, even though the problem is solved in a decomposed form, the decomposed variables are consistent with the original variables at convergence, and the final solution remains feasible for the original problem. This demonstrates that the proposed ADMM decomposition does not introduce any degradation in solution feasibility.

\begin{figure}[!t]
\centering
\includegraphics[width=3.5in]{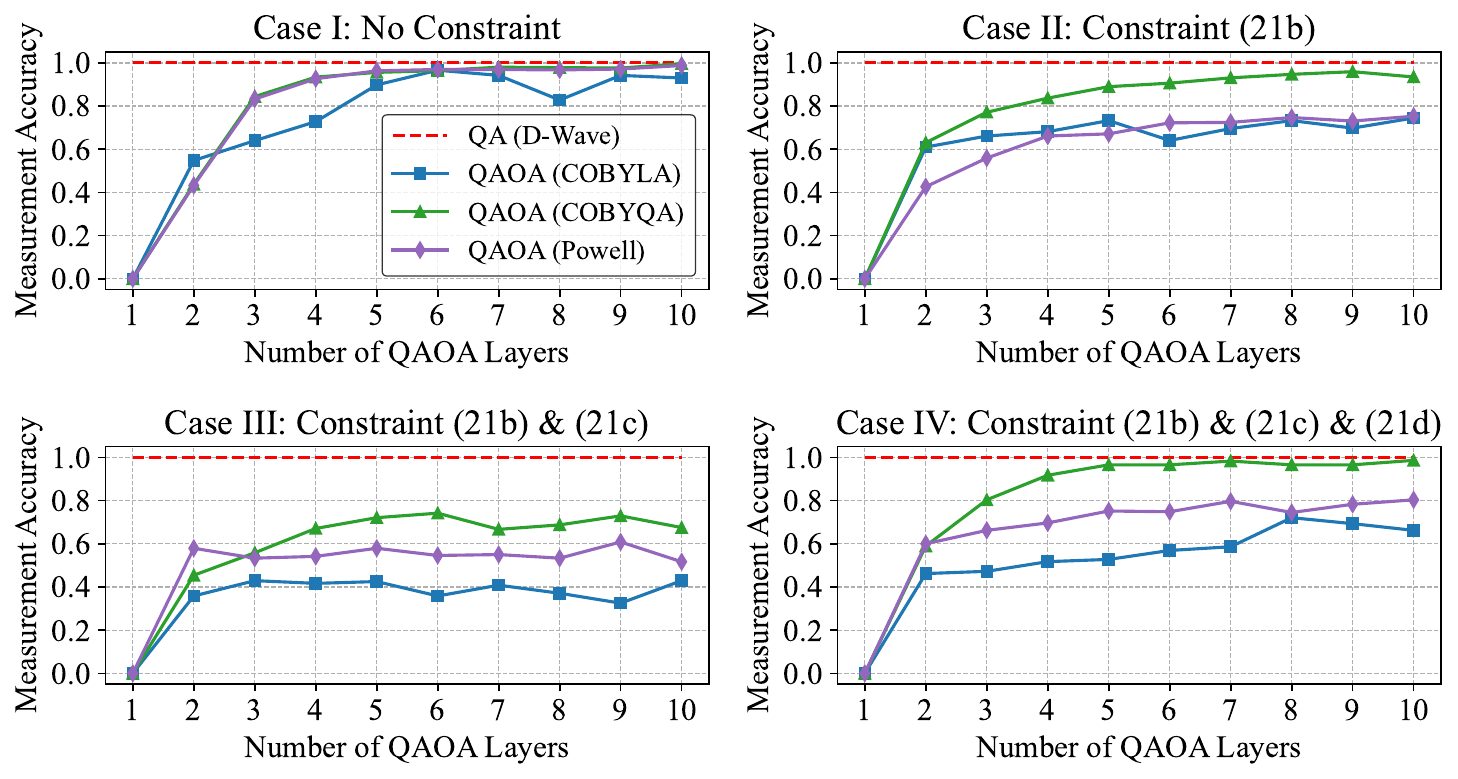}
\vspace{-1.8em}
\caption{QAOA against QA: measurement accuracy under different constraint settings and layer depths.}
\label{fig_QAOA_MA}
\vspace{-0.2em}
\end{figure}

\begin{figure*}[!t]
\centering
\includegraphics[width=6.0in]{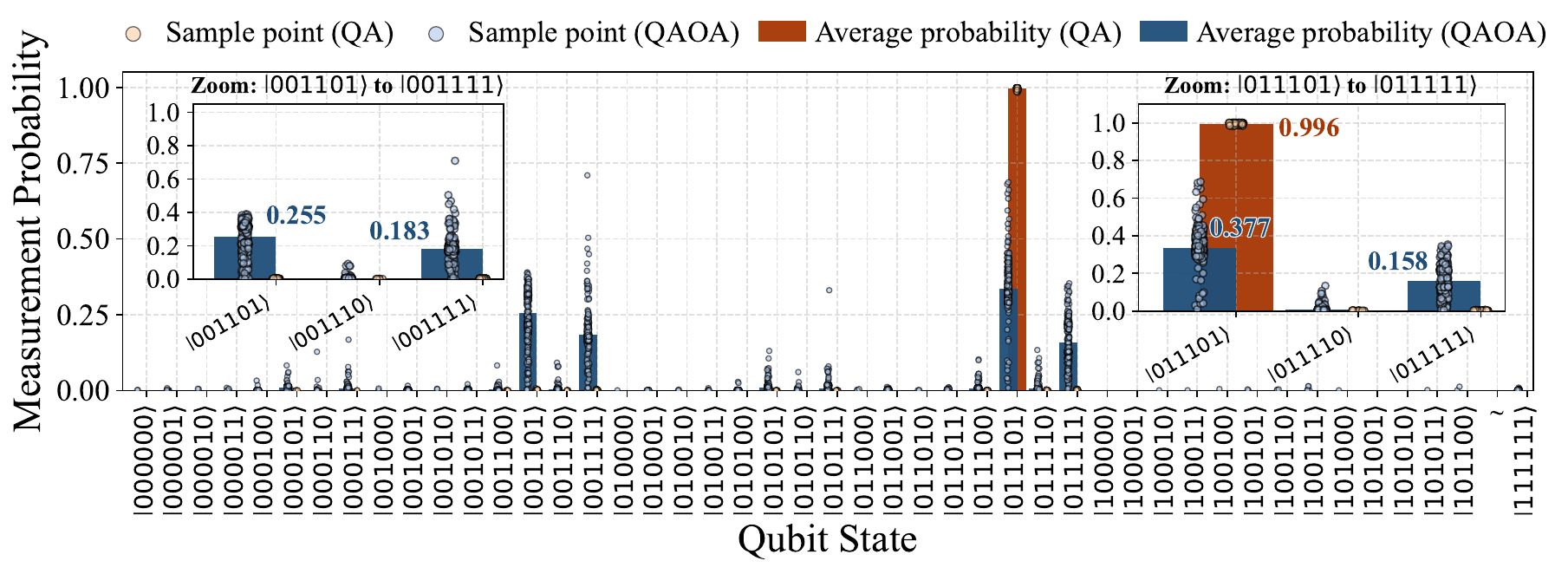}
\vspace{-1.5em}
\caption{QA and QAOA measurement distributions under constraint (\ref{eq:19b}) using QPHR-ADMM.}
\label{fig_QAOA_MP}
\vspace{-1em}
\end{figure*}

At last, QAOA can be integrated into the QPHR-ADMM framework as an alternative QUBO solver. However, as shown in Fig. \ref{fig_QAOA_MA}, its constraint satisfaction is inferior to QA. This is attributed to the dispersed solution distribution of QAOA, illustrated in Fig. \ref{fig_QAOA_MP}, where the probability of the optimal solution is insufficient to suppress suboptimal outcomes. In the BD framework, such inaccuracies may introduce erroneous Benders cuts and impair subsequent iterations. Therefore, QA remains better suited for QPHR-ADMM, as discussed in Section \ref{sec:qaoa_into}.

\begin{table}
\setlength\tabcolsep{3.5pt}
\begin{center}
\caption{The Optimization Results of \eqref{eq:19} Based on QPHR-ADMM.}
\label{tab1}
\renewcommand{\arraystretch}{1.1}
\begin{tabular}{cccccc|c}
\toprule[1.2pt]
\hline
 Constraints& \makecell{Target \\Solutions} & $\sigma^0$ & \makecell{Quantum \\Results} & Iter. & \makecell{Qubit\\ (QPHR\\ADMM)} & \makecell{Qubit\\(Basic\\ QA)}  \\ \hline
 No Constraint &  $001101$ & 0.3 & $\left | 001101  \right \rangle$ & 1 & 2 & 6 \\
 (\ref{eq:19b})&  $011101$ & 0.3 & $\left | 011101  \right \rangle$ & 3 & 2 & 6+$\mathcal{F}$\\
 (\ref{eq:19b}) \& (\ref{eq:19c}) &  $011110$ & 0.5 & $\left | 011110  \right \rangle$& 19 & 2 & 6+$2\mathcal{F}$\\
 (\ref{eq:19b}), (\ref{eq:19c}) \& (\ref{eq:19d})& $110101$ & 0.5 & $\left | 110101  \right \rangle$& 4 & 2 & 6+$3\mathcal{F}$\\ 
\bottomrule[1pt]
\end{tabular}
\end{center}
\vspace{-1.5em}
\end{table}

\subsection{QPU Timing Structure and Runtime Definitions}
\label{app}

This appendix defines the timing components used in the quantum workflow. The definitions follow the Dwave Solver API timing description \cite{dwave_leap_Operation}. Table~\ref{tab:timing_components} summarizes the notation used in the timing analysis.

The sampling time is defined as
\begin{equation}
\setlength\abovedisplayskip{1pt}
\setlength\belowdisplayskip{5pt}
T_{\mathrm{samp}}
=T_{\mathrm{anneal}}+T_{\mathrm{readout}}+T_{\mathrm{delay}} 
\label{eq:sampling_time_appendix}
\end{equation}

The QPU access time reported by the Solver API is given as
\begin{equation}
\setlength\abovedisplayskip{1pt}
\setlength\belowdisplayskip{5pt}
T_{\mathrm{QPU}}=T_{\mathrm{prog}}+T_{\mathrm{samp}}
\label{eq:qpu_access_time_appendix}
\end{equation}

The orchestration time collects the overhead outside the QPU access time and is written as
\begin{equation}
\setlength\abovedisplayskip{1pt}
\setlength\belowdisplayskip{5pt}
T_{\mathrm{orch}}=T_{\mathrm{emb}}+T_{\mathrm{queue}}+
T_{\mathrm{post}}+T_{\mathrm{net}}+T_{\mathrm{other}}
\label{eq:orchestration_time_appendix}
\end{equation}

Therefore, the E2E wall-clock time of the quantum workflow is defined as
\begin{equation}
\setlength\abovedisplayskip{1pt}
\setlength\belowdisplayskip{5pt}
T_{\mathrm{E2E}}=T_{\mathrm{QPU}}+T_{\mathrm{orch}}
\label{eq:end_to_end_time_appendix}
\end{equation}

Because the official Solver API does not separately report all orchestration components, this work reports \(T_{\mathrm{orch}}\) as an aggregated overhead term.

\begin{table}[h]
\centering
\caption{Definition of timing components in the Dwave workflow.}
\label{tab:timing_components}
\begin{tabular}{ll}
\toprule[1.1pt]
\hline
\textbf{Symbol} & \textbf{Description} \\
\midrule
\(T_{\mathrm{samp}}\) & Total QPU sampling time. \\
\(T_{\mathrm{anneal}}\) & Total annealing time over all reads. \\
\(T_{\mathrm{readout}}\) & Total readout time over all reads. \\
\(T_{\mathrm{delay}}\) & Total delay time over all reads. \\
\(T_{\mathrm{prog}}\) & Total QPU programming time. \\
\(T_{\mathrm{QPU}}\) & QPU access time reported by the Solver API. \\
\(T_{\mathrm{orch}}\) & Overhead outside the QPU access time. \\
\(T_{\mathrm{emb}}\) & Minor-embedding construction time. \\
\(T_{\mathrm{queue}}\) & Queue latency before QPU execution. \\
\(T_{\mathrm{post}}\) & Postprocessing overhead time. \\
\(T_{\mathrm{net}}\) & Internet communication latency. \\
\(T_{\mathrm{other}}\) & Other time overheads. \\
\(T_{\mathrm{E2E}}\) & End-to-end wall-clock time of the quantum workflow. \\
\bottomrule
\end{tabular}
\end{table}

\bibliographystyle{IEEEtran}
\bibliography{Reference}

@ARTICLE{teng2016stochastic,
  author={Teng, Fei and Trovato, Vincenzo and Strbac, Goran},
  journal={IEEE Trans. Power Syst.}, 
  title={Stochastic Scheduling With Inertia-Dependent Fast Frequency Response Requirements}, 
  year={2016},
  volume={31},
  number={2},
  pages={1557-1566},
  doi={10.1109/TPWRS.2015.2434837}}

@ARTICLE{bellizio2023transition,
  author={Bellizio, Federica and Xu, Wangkun and Qiu, Dawei and Ye, Yujian and Papadaskalopoulos, Dimitrios and Cremer, Jochen L. and Teng, Fei and Strbac, Goran},
  journal={Proc. IEEE}, 
  title={Transition to Digitalized Paradigms for Security Control and Decentralized Electricity Market}, 
  year={2023},
  volume={111},
  number={7},
  pages={744-761},
  doi={10.1109/JPROC.2022.3161053}}

@ARTICLE{Zhang2017ChanceConstrainedTUC,
  author={Zhang, Yao and Wang, Jianxue and Zeng, Bo and Hu, Zechun},
  journal={IEEE Trans. Power Syst.}, 
  title={Chance-Constrained Two-Stage Unit Commitment Under Uncertain Load and Wind Power Output Using Bilinear Benders Decomposition}, 
  year={2017},
  month={Sept.},
  volume={32},
  number={5},
  pages={3637-3647},
  doi={10.1109/TPWRS.2017.2655078}}

@ARTICLE{Velloso2020TwoStageRUC,
  author={Velloso, Alexandre and Street, Alexandre and Pozo, David and Arroyo, José M. and Cobos, Noemi G.},
  journal={IEEE Trans. Sustain. Energy}, 
  title={Two-Stage Robust Unit Commitment for Co-Optimized Electricity Markets: An Adaptive Data-Driven Approach for Scenario-Based Uncertainty Sets}, 
  year={2020},
  month={Apr.},
  volume={11},
  number={2},
  pages={958-969},
  doi={10.1109/TSTE.2019.2915049}}

@article{Huang2014TwostageSUC,
title = {Two-stage stochastic unit commitment model including non-generation resources with conditional value-at-risk constraints},
journal = {Electr. Power Syst. Res.},
volume = {116},
pages = {427-438},
year = {2014},
month={Nov.},
issn = {0378-7796},
doi = {https://doi.org/10.1016/j.epsr.2014.07.010},
author = {Yuping Huang and Qipeng P. Zheng and Jianhui Wang}
}

@ARTICLE{Asensio2016StochasticUC,
  author={Asensio, Miguel and Contreras, Javier},
  journal={IEEE Trans. Smart Grid}, 
  title={Stochastic Unit Commitment in Isolated Systems With Renewable Penetration Under CVaR Assessment}, 
  year={2016},
  month={May},
  volume={7},
  number={3},
  pages={1356-1367},
  doi={10.1109/TSG.2015.2469134}}

@article{Xiong2023MultistageRD,
  title={Multi-stage robust dynamic unit commitment based on pre-extended-fast robust dual dynamic programming},
  author={Xiong, HouBo and Shi, YunHui and Chen, Zhe and Guo, Chuangxin and Ding, Yi},
  journal={IEEE Trans. Power Syst.},
  volume={38},
  number={3},
  pages={2411--2422},
  year={2023},
  month={May},
  publisher={IEEE}
}

@ARTICLE{Nasri2016NetworkConstrainedAC,
  author={Nasri, Amin and Kazempour, S. Jalal and Conejo, Antonio J. and Ghandhari, Mehrdad},
  journal={IEEE Trans. Power Syst.}, 
  title={Network-Constrained AC Unit Commitment Under Uncertainty: A Benders’ Decomposition Approach}, 
  year={2016},
  month={Jan.},
  volume={31},
  number={1},
  pages={412-422},
  doi={10.1109/TPWRS.2015.2409198}}

@article{Colonetti2020CombiningLR,
  title={Combining Lagrangian relaxation, benders decomposition, and the level bundle method in the stochastic hydrothermal unit-commitment problem},
  author={Colonetti, Bruno and Finardi, Erlon C},
  journal={Int. Trans. Electr. Energy Syst.},
  volume={30},
  number={9},
  pages={1-22},
  year={2020},
  month={Sept.},
  publisher={Wiley Online Library}
}

@book{Nielsen2010QC,
  title={Quantum computation and quantum information},
  author={Nielsen, Michael A and Chuang, Isaac L},
  year={2010},
  publisher={Cambridge university press}
}

@article{Morstyn2024OpportunitiesQC,
  title={Opportunities for quantum computing within net-zero power system optimization},
  author={Morstyn, Thomas and Wang, Xiangyue},
  journal={Joule},
  year={2024},
  month={Jun.},
  volume={8},
  number={6},
  pages={1619-1640},
  publisher={Elsevier}
}

@inproceedings{Amani2023QuantumenhancedDC,
  title={Quantum-enhanced DC optimal power flow},
  author={Amani, Farshad and Mahroo, Reza and Kargarian, Amin},
  booktitle={2023 IEEE Texas Power Energy Conf.},
  pages={1--6},
  year={2023},
}

@article{Feng2021QuantumPF,
  title={Quantum power flow},
  author={Feng, Fei and Zhou, Yifan and Zhang, Peng},
  journal={IEEE Trans. Power Syst.},
  volume={36},
  number={4},
  pages={3810--3812},
  year={2021},
  month={Jul.},
  publisher={IEEE}
}

@article{Nikmehr2023QuantuminspiredPS,
  title={Quantum-inspired power system reliability assessment},
  author={Nikmehr, Nima and Zhang, Peng},
  journal={IEEE Trans. Power Syst.},
  volume={38},
  number={4},
  pages={3476--3490},
  year={2023},
  month={Jul.},
  publisher={IEEE}
}

@article{Mahroo2023LearningIQ,
  title={Learning infused quantum-classical distributed optimization technique for power generation scheduling},
  author={Mahroo, Reza and Kargarian, Amin},
  journal={IEEE Trans. Quantum Eng.},
  year={2023},
  month={Nov.},
  volume={4},
  pages={1-14},
  publisher={IEEE}
}

@article{Nikmehr2022QuantumDUC,
  title={Quantum distributed unit commitment: An application in microgrids},
  author={Nikmehr, Nima and Zhang, Peng and Bragin, Mikhail A},
  journal={IEEE Trans. Power Syst.},
  volume={37},
  number={5},
  pages={3592--3603},
  year={2022},
  month={Sept.},
  publisher={IEEE}
}

@article{Feng2023NovelRUC,
  title={Novel resolution of unit commitment problems through quantum surrogate Lagrangian relaxation},
  author={Feng, Fei and Zhang, Peng and Bragin, Mikhail A and Zhou, Yifan},
  journal={IEEE Trans. Power Syst.},
  volume={38},
  number={3},
  pages={2460--2471},
  year={2023},
  month={May},
  publisher={IEEE}
}

@article{Pastorello2019QuantumAL,
  title={Quantum annealing learning search for solving QUBO problems},
  author={Pastorello, Davide and Blanzieri, Enrico},
  journal={Quantum Inf. Process.},
  volume={18},
  number={10},
  pages={303},
  year={2019},
  month={Aug.},
  publisher={Springer}
}

@ARTICLE{Wang2023QuantumAIS,
  author={Wang, Dawei and Zheng, Kedi and Teng, Fei and Chen, Qixin},
  journal={IEEE Trans. Power Syst.}, 
  title={Quantum Annealing With Integer Slack Variables for Grid Partitioning}, 
  year={2023},
  month={Mar.},
  volume={38},
  number={2},
  pages={1747-1750},
  doi={10.1109/TPWRS.2022.3229862}}

@article{Morstyn2023AnnealingbasedQC,
  title={Annealing-based quantum computing for combinatorial optimal power flow},
  author={Morstyn, Thomas},
  journal={IEEE Trans. Smart Grid},
  volume={14},
  number={2},
  pages={1093--1102},
  year={2023},
  month={Mar.},
  publisher={IEEE}
}

@article{Gao2022HybridQC,
  title={Hybrid quantum-classical general benders decomposition algorithm for unit commitment with multiple networked microgrids},
  author={Gao, Fang and Huang, Dejian and Zhao, Ziwei and Dai, Wei and Yang, Mingyu and Shuang, Feng},
  journal={arXiv preprint arXiv:2210.06678},
  year={2022}
}

@article{Leenders2024IntegratingQCC,
  title={Integrating quantum and classical computing for multi-energy system optimization using Benders decomposition},
  author={Leenders, Ludger and Sollich, Martin and Reinert, Christiane and Bardow, Andr{\'e}},
  journal={Comput. Chem. Eng.},
  pages={108763},
  year={2024},
  publisher={Elsevier}
}

@article{Roald2023PowerSO,
  title={Power systems optimization under uncertainty: A review of methods and applications},
  author={Roald, Line A and Pozo, David and Papavasiliou, Anthony and Molzahn, Daniel K and Kazempour, Jalal and Conejo, Antonio},
  journal={Electr. Power Syst. Res.},
  volume={214},
  pages={108725},
  year={2023},
  month={Jan.},
  publisher={Elsevier}
}

@article{Albash2018AdiabaticQC,
  title={Adiabatic quantum computation},
  author={Albash, Tameem and Lidar, Daniel A},
  journal={Rev. Mod. Phys.},
  volume={90},
  number={1},
  pages={015002},
  year={2018},
  month={Jan.},
  publisher={APS}
}

@manual{DWave2024QPUSpecificPP,
  title        = {QPU-Specific Physical Properties: Advantage\_system4.1},
  year         = 2024,
  address      = {Burnaby, BC, CA},
  organization = {D-Wave Systems Inc.},
}

@article{Rockafellar1974AugmentedLM,
  title={Augmented Lagrange multiplier functions and duality in nonconvex programming},
  author={Rockafellar, R Tyrrell},
  journal={SIAM J. Contr.},
  volume={12},
  number={2},
  pages={268--285},
  year={1974},
  month={May},
  publisher={SIAM}
}

@article{Han2012noteADMM,
  title={A note on the alternating direction method of multipliers},
  author={Han, Deren and Yuan, Xiaoming},
  journal={J. Optim. Theory Appl.},
  volume={155},
  pages={227--238},
  year={2012},
  month={Feb.},
  publisher={Springer}
}

@article{Hong2017linearCADMM,
  title={On the linear convergence of the alternating direction method of multipliers},
  author={Hong, Mingyi and Luo, Zhi Quan},
  journal={Math. Program.},
  volume={162},
  number={1},
  pages={165--199},
  year={2017},
  month={Mar.},
  publisher={Springer}
}

@INPROCEEDINGS{Zhao2021HybridQB,
  author={Zhao, Zhongqi and Fan, Lei and Han, Zhu},
  booktitle={2022 IEEE Wireless Commun. Netw. Conf.}, 
  title={Hybrid Quantum Benders’ Decomposition For Mixed-integer Linear Programming}, 
  year={2022},
  volume={},
  number={},
  pages={2536-2540},
  doi={10.1109/WCNC51071.2022.9771632}}

@ARTICLE{Lorca2017MultistageRUC,
  author={Alvaro Lorca and Xu Andy Sun},
  journal={IEEE Trans. Power Syst.}, 
  title={Multistage Robust Unit Commitment With Dynamic Uncertainty Sets and Energy Storage}, 
  year={2017},
  volume={32},
  number={3},
  month={May},
  pages={1678-1688},
  doi={10.1109/TPWRS.2016.2593422}}

@article{FU2023CoordinatedPR,
title = {Coordinated post-disaster restoration for resilient urban distribution systems: A hybrid quantum-classical approach},
journal = {Energy},
volume = {284},
pages = {129314},
year = {2023},
  month={Dec.},
issn = {0360-5442},
doi = {https://doi.org/10.1016/j.energy.2023.129314},
author = {Wei Fu and Haipeng Xie and Hao Zhu and Hefeng Wang and Lizhou Jiang and Chen Chen and Zhaohong Bie},

}

@article{Elijah2023ComparingTG,
  title={Comparing Three Generations of D-Wave Quantum Annealers for Minor Embedded Combinatorial Optimization Problems},
  author={Elijah Pelofske},
  journal={arXiv preprint arXiv:2301.03009},
  year={2023}
}

@ARTICLE{Kimleang2023LeveragingKQAOA,
  author={Kea, Kimleang and Huot, Chansreynich and Han, Youngsun},
  journal={IEEE Access}, 
  title={Leveraging Knapsack QAOA Approach for Optimal Electric Vehicle Charging}, 
  year={2023},
  volume={11},
  number={},
  pages={109964-109973},
  doi={10.1109/ACCESS.2023.3320800}}

@ARTICLE{Maneesh2020AnOptimalMVCNL,
  author={Kumar, Maneesh and Tyagi, Barjeev},
  journal={IEEE Syst. J.}, 
  title={An Optimal Multivariable Constrained Nonlinear (MVCNL) Stochastic Microgrid Planning and Operation Problem With Renewable Penetration}, 
  year={2020},
  volume={14},
  number={3},
  pages={4143-4154},
  doi={10.1109/JSYST.2019.2963729}}

@ARTICLE{Pablo2009UncertaintyMUC,
  author={Ruiz, Pablo A. and Philbrick, C. Russ and Zak, Eugene and Cheung, Kwok W. and Sauer, Peter W.},
  journal={IEEE Trans. Power Syst.}, 
  title={Uncertainty Management in the Unit Commitment Problem}, 
  year={2009},
  volume={24},
  number={2},
  pages={642-651},
  doi={10.1109/TPWRS.2008.2012180}}

@ARTICLE{Ignacio2017AnEfficientRS,
  author={Blanco, Ignacio and Morales, Juan M.},
  journal={IEEE Trans. Power Syst.}, 
  title={An Efficient Robust Solution to the Two-Stage Stochastic Unit Commitment Problem}, 
  year={2017},
  volume={32},
  number={6},
  pages={4477-4488},
  doi={10.1109/TPWRS.2017.2683263}}

@ARTICLE{Canan2016AnImprovedSUC,
  author={Uckun, Canan and Botterud, Audun and Birge, John R.},
  journal={IEEE Trans. Power Syst.}, 
  title={An Improved Stochastic Unit Commitment Formulation to Accommodate Wind Uncertainty}, 
  year={2016},
  volume={31},
  number={4},
  pages={2507-2517},
  doi={10.1109/TPWRS.2015.2461014}}

@article{Farhi2014AQuantumAOA,
  title={A Quantum Approximate Optimization Algorithm},
  author={Farhi, Edward and Goldstone, Jeffrey and Gutmann, Sam},
  journal={arXiv preprint arXiv:1411.4028},
  year={2014}
}

@ARTICLE{Blekos2024AreviewQAOA,
  author = {Kostas Blekos and Dean Brand and Andrea Ceschini and Chiao-Hui Chou and Rui-Hao Li and Komal Pandya and Alessandro Summer},
  journal={Phys. Rep.}, 
  title={An Improved Stochastic Unit Commitment Formulation to Accommodate Wind Uncertainty}, 
  year={2024},
  volume = {1068},
  pages = {1-66},
  doi={https://doi.org/10.1016/j.physrep.2024.03.002},
}

@article{Kim2024DistributedQAOA,
  title={Distributed Quantum Approximate Optimization Algorithm on Integrated High-Performance Computing and Quantum Computing Systems for Large-Scale Optimization},
  author={Kim, Seongmin and Luo, Tengfei and Lee, Eungkyu and Suh, In-Saeng},
  journal={arXiv preprint arXiv: 2407.20212},
  year={2024}
}

@INPROCEEDINGS{Pelofske2023QuantumAnnealingvsQAOA,
  author={Pelofske, Elijah
and Bartschi, Andreas
and Eidenbenz, Stephan},
  booktitle={Int. Conf. High Perform. Comput.}, 
  title={Quantum Annealing vs. QAOA: 127 Qubit Higher-Order Ising Problems on NISQ Computers}, 
  year={2023},
  volume={},
  number={},
  pages={240-258},
}

@INPROCEEDINGS{Halffmann2022QCApproach,
  author={Pascal Halffmann and  Patrick Holzer and Kai Plociennik and  Michael Trebing},
  booktitle={Int. Conf. Oper. Res.}, 
  title={A Quantum Computing Approach for the Unit
Commitment Problem}, 
  year={2022},
  volume={},
  number={},
  pages={113-120},
}

@article{Paterakis2023hybridQC,
  title={Hybrid quantum-classical multi-cut benders approach with a power system application},
  author={Paterakis, Nikolaos G},
  journal={Comput. Chem. Eng.},
  volume={172},
  pages={108161},
  year={2023},
  publisher={Elsevier}
}

@article{Ling2025hybridQA,
  title={Hybrid quantum annealing decomposition framework for unit commitment},
  author={Ling, Jiajie and Zhang, Quan and Geng, Guangchao and Jiang, Quanyuan},
  journal={Electr. Power Syst. Res.},
  volume={238},
  pages={111121},
  year={2025},
  publisher={Elsevier}
}

@article{wang2021noise,
  title={Noise-induced barren plateaus in variational quantum algorithms},
  author={Wang, Samson and Fontana, Enrico and Cerezo, Marco and Sharma, Kunal and Sone, Akira and Cincio, Lukasz and Coles, Patrick J},
  journal={Nat. Commun.},
  volume={12},
  number={1},
  pages={6961},
  year={2021},
  publisher={Nature Publishing Group UK London}
}

@article{willsch2022benchmarking,
  title={Benchmarking Advantage and D-Wave 2000Q quantum annealers with exact cover problems},
  author={Willsch, Dennis and Willsch, Madita and Calaza, Carlos D Gonzalez and Jin, Fengping and De Raedt, Hans and Svensson, Marika and Michielsen, Kristel},
  journal={Quantum Inf. Process.},
  volume={21},
  number={4},
  year={2022},
  publisher={Springer Science and Business Media LLC}
}

@inproceedings{barrass2025leveraging,
  title={Leveraging Quantum Computing for Accelerated Classical Algorithms in Power Systems Optimization},
  author={Barrass, Rosemary and Nagarajan, Harsha and Coffrin, Carleton},
  booktitle={Proc. Int. Conf. Integr. Constraint Program., Artif. Intell., Oper. Res.},
  pages={52--67},
  year={2025},
  organization={Springer}
}

@article{gao2025distributed,
  title={Distributed quantum generalized benders decomposition for unit commitment problems: F. Gao et al.},
  author={Gao, Fang and Huang, Dejian and Zhao, Ziwei and Dai, Wei and Yang, Mingyu and Gao, Qing and Pan, Yu},
  journal={Quantum Inf. Process.},
  volume={24},
  number={12},
  pages={376},
  year={2025},
  publisher={Springer}
}

@article{zhao2023optimal,
  title={Optimal data center energy management with hybrid quantum-classical multi-cuts benders' decomposition method},
  author={Zhao, Zhongqi and Fan, Lei and Han, Zhu},
  journal={IEEE Trans. Sustain. Energy},
  volume={15},
  number={2},
  pages={847--858},
  year={2023},
  publisher={IEEE}
}

@article{wei2024hybrid,
  title={Hybrid quantum-classical benders' decomposition for federated learning scheduling in distributed networks},
  author={Wei, Xinliang and Fan, Lei and Guo, Yuanxiong and Gong, Yanmin and Han, Zhu and Wang, Yu},
  journal={IEEE Trans. Netw. Sci. Eng.},
  year={2024},
  publisher={IEEE}
}

@article{geoffrion1972generalized,
  title={Generalized benders decomposition},
  author={Geoffrion, Arthur M},
  journal={J. Optim. Theory Appl.},
  volume={10},
  number={4},
  pages={237--260},
  year={1972},
  publisher={Springer}
}

@article{wang2024customized,
  title={A Customized Augmented Lagrangian Method for Block-Structured Integer Programming},
  author={Wang, Rui and Zhang, Chuwen and Pu, Shanwen and Gao, Jianjun and Wen, Zaiwen},
  journal={IEEE Trans. Pattern Anal. Mach. Intell.},
  volume={46},
  number={12},
  pages={9439--9455},
  year={2024},
  publisher={IEEE}
}

@article{wuijts2023new,
  title={New efficient ADMM algorithm for the Unit Commitment Problem},
  author={Wuijts, Rogier Hans and Akker, Marjan van den and Broek, Machteld van den},
  journal={arXiv preprint arXiv:2311.13438},
  year={2023}
}

@article{mhanna2018adaptive,
  title={Adaptive ADMM for distributed AC optimal power flow},
  author={Mhanna, Sleiman and Verbi{\v{c}}, Gregor and Chapman, Archie C},
  journal={IEEE Trans. Power Syst.},
  volume={34},
  number={3},
  pages={2025--2035},
  year={2018},
  publisher={IEEE}
}

@article{pelofske2025comparing,
  title={Comparing three generations of D-Wave quantum annealers for minor embedded combinatorial optimization problems},
  author={Pelofske, Elijah},
  journal={Quantum Sci. Technol.},
  volume={10},
  number={2},
  pages={025025},
  year={2025},
  publisher={IOP Publishing}
}

@misc{dwave_leap_Operation,
  title        = {Operation and Timing},
  author       = {{D-Wave Quantum Inc.}},
  year         = {2025},
  howpublished = {\url{https://docs.dwavequantum.com/en/latest/quantum_research/operation_timing.html}},

}

\newpage

\vfill

\end{document}